\documentclass{aa}  

\usepackage{graphicx}

\usepackage{txfonts}
\usepackage{placeins}

\usepackage{natbib}
\bibpunct{(}{)}{;}{a}{}{,}

\usepackage[colorlinks=true, urlcolor=blue, colorlinks=true, citecolor=blue, linkcolor=blue]{hyperref}

\begin{document} 
   
   \title{XMAGNET: Velocity structure functions of active galactic nucleus-driven turbulence in the multiphase intracluster medium}
   
\author{M. Fournier\inst{1}, P. Grete\inst{1}, M. Brüggen\inst{1},  
           B. W. O'Shea\inst{2,3,4,5},
           D. Prasad\inst{6}, \\
           B. D. Wibking\inst{3},  
           F. W. Glines\inst{7},  
           R. Mohapatra\inst{8}
          }

\institute{Universität Hamburg, Hamburger Sternwarte, Gojenbergsweg 112, 21029 Hamburg, Germany \\
              \email{martin.fournier@uni-hamburg.de}
        \and
    Department of Computational Mathematics, Science, and Engineering, Michigan State University, East Lansing, MI 48824, USA
         \and
    Department of Physics and Astronomy, Michigan State University, East Lansing, MI 48824, USA
        \and
    Facility for Rare Isotope Beams, Michigan State University, East Lansing, MI 48824, USA
        \and
    Institute for Cyber-Enabled Research, 567 Wilson Road, Michigan State University, East Lansing, MI 48824
        \and
    School of Physics and Astronomy, Cardiff University, 5 The Parade, Cardiff CF24 3AA, UK
        \and
    Theoretical Division, TA-3 Bldg. 123, Los Alamos National Laboratory, Los Alamos, NM 87545 
        \and
    Department of Astrophysical Sciences, Princeton University, NJ 08544, USA
}

   \date{Received \today}

  \abstract
   {Significant theoretical and observational efforts are underway to investigate the properties of the turbulence in the hot plasma that pervades galaxy clusters. Spectroscopy has been used to study the projected line-of-sight velocities in both the hot intracluster medium and the cold gas phase in combination with optical and X-ray telescopes.}
   {In this work, we characterize the velocity structure functions (VSFs) of the multiphase intracluster medium in a simulated galaxy cluster core and study the effects of projections on the hot and cold phase of the gas.}
   {We used the fiducial run of the XMAGNET suite, a collection of exascale magneto-hydrodynamical simulations of a cool-core cluster, to compute VSFs. The simulation includes radiative cooling as well as a model for active galactic nuclei feedback.}
   {Examining three-dimensional and line-of-sight VSFs, we find no clear correlation between the behavior of the hot ($10^6\, \mathrm{K}\, \leq T \leq 10^8 \, \mathrm{K}$) and cold ($T\leq 10^5$ K) phase VSFs. Assuming a power-law model for the VSF, we find that the power-law index $m$ of the cold phase varies significantly throughout the 4 Gyr simulation time. We compared our VSFs with observations using mock optical and X-ray images, and we conclude that projection effects significantly impact the amplitude and power-law index of both the hot and cold phases. In the cold phase, applying a Gaussian smoothing filter to model effects of atmospheric seeing significantly increases the  index of the projected VSF at scales below the filter's kernel size. Moreover, the VSF amplitude and power-law index vary significantly depending on the viewing orientation.
   }
   {Observational biases such as projection effects, atmospheric seeing, and the viewing angle cannot be ignored when interpreting the line-of-sight velocity structure of the intracluster medium.}
   \titlerunning{XMAGNET: Velocity structure of AGN-driven turbulence}
   \authorrunning{M. Fournier et al.} 
   \maketitle

\setlength{\parindent}{0cm}

\section{Introduction}

Around 10\% of the total mass of galaxy clusters is thought to be contained in a hot plasma of ionized gas, usually designated as the intracluster medium \citep[ICM; e.g.,][]{Chiu_2015}. Cooling via Bremsstrahlung gives rise to the emission of radiations and makes clusters strong X-ray sources \citep{Felten_1966,Mitchell_1976,Serlemitsos_1977,Sarazin_1986}. Mass estimates of clusters, as well as density or temperature profiles, can be derived from observations of this emission, assuming that the ICM is in hydrostatic equilibrium. However, both simulations and observations have shown that several processes are likely to cause deviations to the hydrostatic equilibrium \citep{Rasia_2006,Lau_2013,Biffi_2016,Barnes_2021}. In particular, it is expected that non-thermal sources of pressure, such as turbulence, might add to the thermal pressure support of the ICM against gravity \citep{Fusco_Femiano_2017,Angelinelli_2020,Vazza_2011}. 

Cluster mergers are thought to be the main drivers of turbulence on the scale of entire clusters. In a typical merger event, two clusters collide with relative velocities of $\sim 10^3 \, \mathrm{km}\,\mathrm{s}^{-1}$. As the ICM relaxes in the post-merger phase, shocks form and propagate through the ICM, carrying an energy of up to $10^{65}$ erg and providing heat to the plasma on scales of up to tens of Mpc \citep[e.g.,][]{Markevitch_2007}. The subsequent dissipation of turbulence heats the ICM on longer timescales. Results from hydrodynamical simulations suggest that turbulent kinetic energy in perturbed clusters resulting from mergers can account for up to 20 -- 30 \% of the ICM's thermal energy \citep{Dolag_2005,Vazza_2011}. A common consequence of mergers in galaxy clusters is a temporary offset of the gas core with respects to its dark matter counterparts. In such cases, the ICM falls back toward the center of the cluster, and the resulting gas sloshing can drive turbulence on scales of up to 100 kpc \citep{Zuhone_2013}. In the innermost regions of galaxy clusters, active galactic nuclei (AGNs) are suspected to be the main drivers of turbulence \citep{Scannapieco_2008,Li_2020}. By accelerating jets of gas up to relativistic speeds, AGNs inflate cavities in the ICM, which rise buoyantly and propagate out to distances of a few to hundreds of kiloparsecs \citep{Birzan_2004,Dunn_2005,Prunier_2025}. As they rise, these cavities displace the ICM, mix with the surrounding gas, and inject turbulent energy \citep{Bruggen_2009, 2024A&A...692A.108G}. 
Constraining turbulence with observations is challenging and typically relies on indirect measurements. X-ray surface brightness (SB) fluctuations can be computed by subtracting smoothed models from the SB data \citep{Churazov_2012,Zhuravleva_2014}. The velocity power spectrum of the gas can then be inferred from the power spectrum of the SB fluctuations, but it is subject to a range of assumptions. X-ray calorimeters can also provide measurements of the typical turbulent velocities in cluster cores through line broadening measurements \citep{Hitomi,Xrism_2025}. More recently, \citet{Li_2020} have proposed using observations of ionized filaments found in the inner tens of kiloparsecs of cool-core clusters to probe the turbulent motion of the ICM. Such filamentary structures are common in cool-core clusters \citep{Olivares_2019} and are thought to originate from thermal instabilities in the ICM \citep{McCourt_2012}, likely seeded by AGN outflows \citep{Yuan_2014b,Fournier_2024b}. The line-of-sight velocity of these filaments can be obtained from optical spectrometers with angular resolutions of less than $\sim 10^{-1} \, \mathrm{arcsec}^2$. By measuring the velocity structure function (VSF) of the filaments, \citet{Li_2020} found deviations from a Kolmogorov scaling characterized by a steeper power-law index than the expected $\ell^{1/3}$, where $\ell$ designates the spatial scale. The origin of this steepening is debated in the literature. One hypothesis is that it is related to the injection of turbulence caused by the central AGN. As the bubbles rise in the ICM, they could drive turbulence and increase the amplitude of the VSF at scales on the order of their typical size (i.e., $\sim$ 10\,kpc; \citealt{Hillel_2020}). Other possibilities include the contribution of gravitational acceleration \citep{Wang_2021} or the condensation of cold gas within supersonic outflows \citep{Hu_2022}, which have both been found to result in VSFs steeper than the Kolmogorov scaling law. The interpretation of these observations could be affected by projection effects or limited angular resolution. Some of these effects have been investigated in turbulent box simulations \citep{Mohapatra_2022} and idealized cluster setups \citep{2016ApJ...817..110Z,Wang_2021,Sotira_2024}.

\begin{figure*}[h!]
\includegraphics[width=\textwidth]{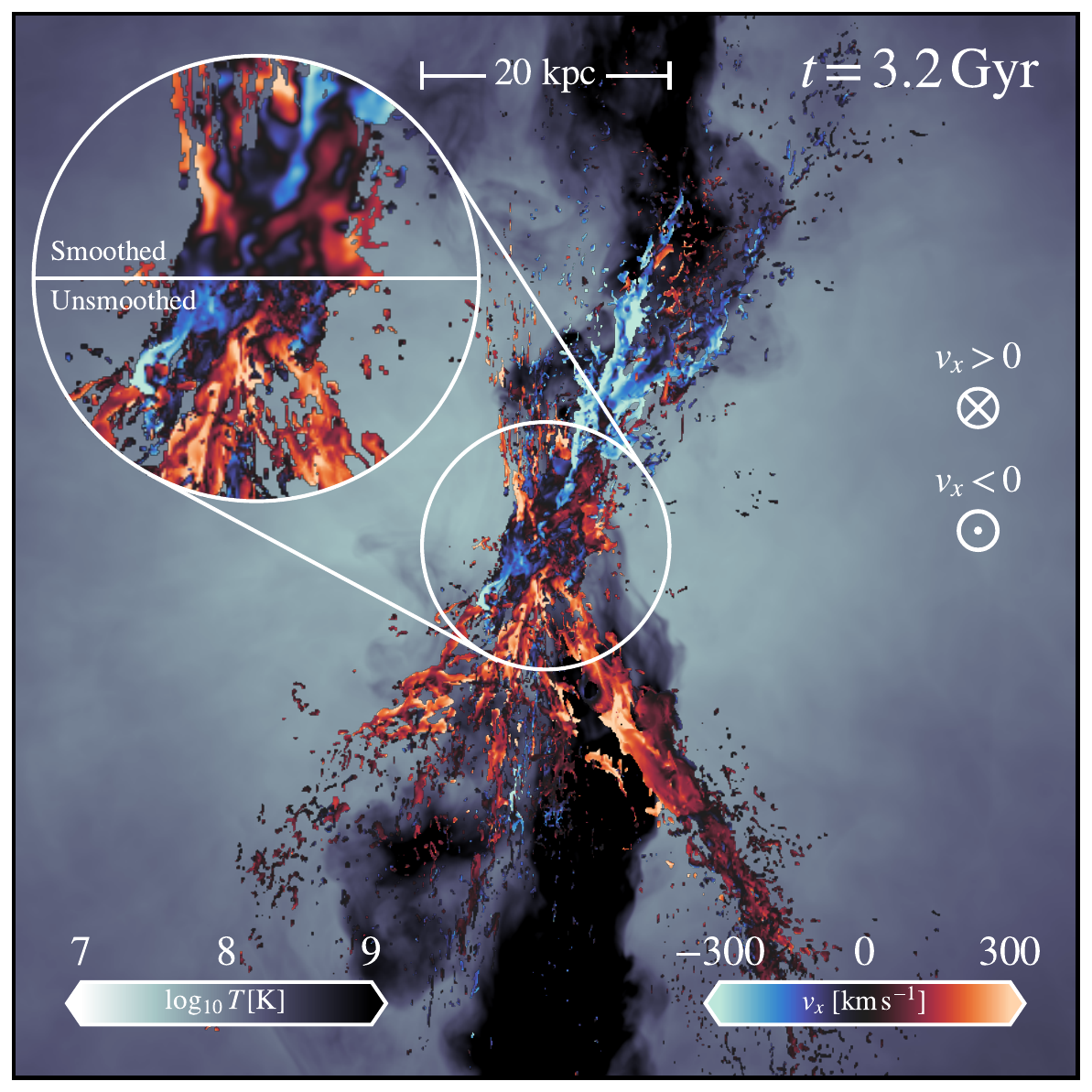}
    \caption{H$\alpha$ emission -- weighted projection of the line-of-sight velocity for the inner 80 kpc of our simulated box at $t=3.2$ Gyr. A zoom-in on the innermost region showcases the effect of atmospheric seeing on the map. The background image is a projection of the temperature field. The figure can be compared to Fig. 2 from \citet{Li_2020}. $\odot$ and $\otimes$ markers indicate the directions pointing toward (away) from the viewer, associated  with negative (positive) line-of-sight velocity (respectively). More detail on the emissivity weighting are provided in Sect.~\ref{sect:Halpha}. A movie showing the three dimensional structure of the cold phase is available as a supplementary material.}
    \label{fig:Li2020}
\end{figure*}

\begin{figure*}[h!]
    \includegraphics[width=\textwidth]{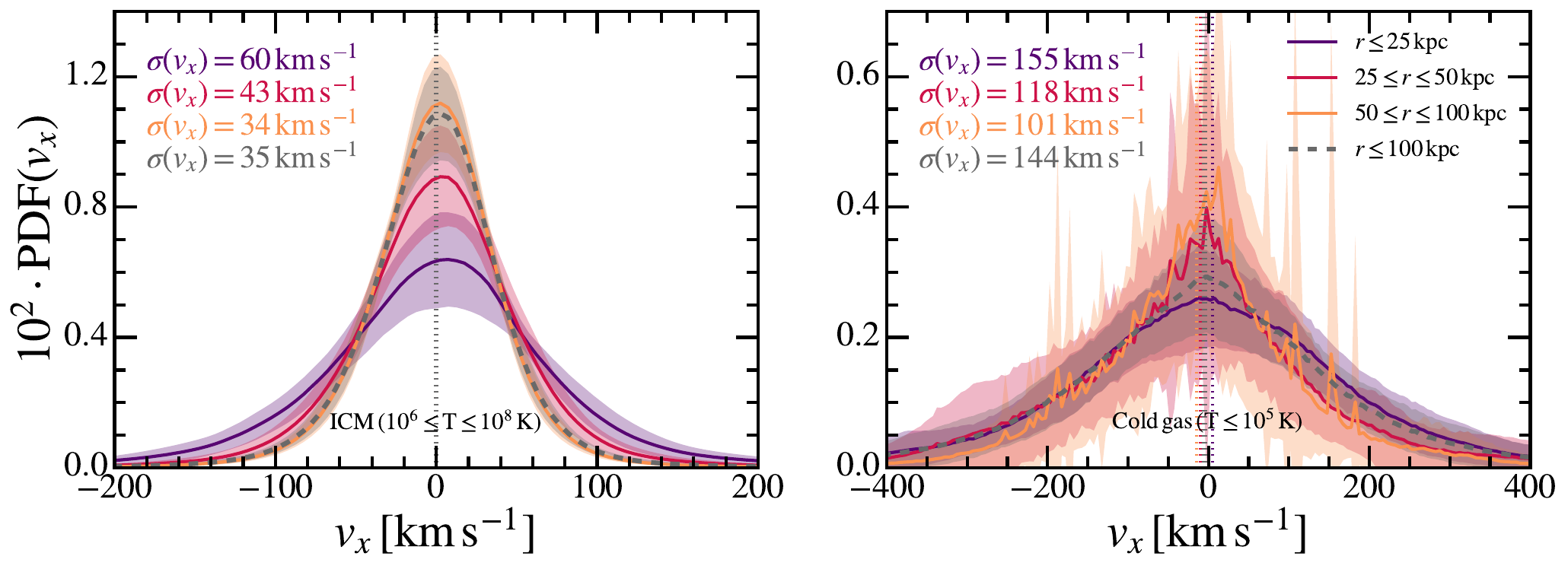}
    \caption{Probability density function (PDF) of the velocities' $x$-component for the hot phase ($10^6 \leq T \leq 10^8 \, \mathrm{K}$, left panel) and cold phase ($T \leq 10^5 \, \mathrm{K}$, right panel) for various radius bins and the PDF of all gas located at radii $r \leq 100$ kpc, averaged between 0.9 and 4 Gyr. Vertical dashed lines indicate the median values. Colored areas represent the standard deviation resulting from time variation. We have verified that the $y$-component leads to similar results.}
    \label{fig:velocitydisp}
\end{figure*}

\begin{figure*}[h!]
    \includegraphics[width=\textwidth]{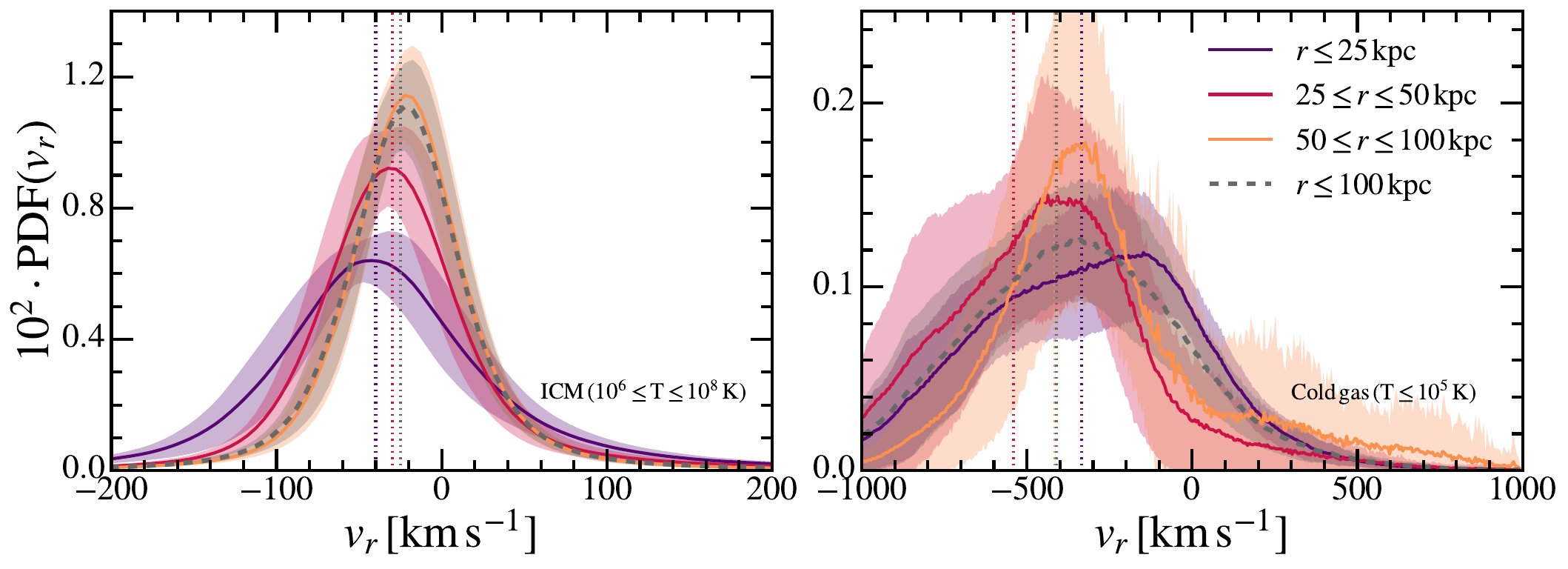}
    \caption{Same as Fig.~\ref{fig:velocitydisp} but for radial velocity, $v_r$.}
    \label{fig:radveldistrib}
\end{figure*}

In this paper, we use a high-resolution exascale magneto-hydrodynamical (MHD) simulation from the XMAGNET project\footnote{See \url{https://xmagnet-simulations.github.io/} for an overview.} to characterize the VSFs of the multi-phase ICM. We explore the effects of projection, emission weighting, atmospheric seeing, and viewing angle on the VSF. In Sect.~\ref{sect:method}, we describe our numerical method and our simulation parameters. In Sect.~\ref{sect:quantify}, we quantify the turbulent motions of both the hot and cold phases using a first-order VSF. In Sect.~\ref{sect:projection}, we investigate up to what extent projection effects might affect the measurement of the VSF before discussing our results in Sect.~\ref{sect:discussion} and concluding in Sect.~\ref{sect:conclusion}.

\section{Method}
\label{sect:method}

\subsection{Simulation suite and initial conditions}

The XMAGNET simulation suite used in this study is introduced and described in \citet{xmagnet_overview}. Hereafter, we summarize the key points of the methodology presented in that paper. 

We ran hydrodynamical and MHD simulations of an idealized cool-core cluster  on the Frontier supercomputer\footnote{Frontier is operated by the Oak Ridge National Laboratory's Leadership Computing Facility on behalf of the Department of Energy. This work is supported by the DOE INCITE program under allocation AST-146 (2023-2024).}. We used {\textsc{AthenaPK}}\footnote{\textsc{AthenaPK} is available at \url{https://github.com/parthenon-hpc-lab/athenapk} and commit \texttt{3ce0a88} was used for the simulations.}, a performance-portable version of the {\scshape{Athena++}} astrophysical magnetohydrodynamics code \citep{Stone_2020}. It is based on the block-structured refinement framework {\scshape{Parthenon}} \citep{Parthenon} and includes magnetic fields, radiative cooling, and AGN and stellar feedback. 
The simulations used an overall second-order accurate, shock-capturing, finite-volume scheme consisting of 
second-order Runge Kutta (RK2) time integration, piecewise-linear reconstruction, and an HLLD Riemann solver \citep{Miyoshi_2005}. 

Gravity is treated as an acceleration term parametrized by a superposition of a Navarro-Frenk-White (NFW) profile \citep{Navarro_1997}, a Hernquist potential representing the brightest central galaxy's mass \citep{Hernquist_1990}, and a central point mass modeling the effect of 3C 84, the supermassive black hole (SMBH) in the Perseus cluster's BCG. The initial gas distribution is calculated numerically to obey hydrostatic equilibrium assuming the entropy profile of the Perseus cluster from the ACCEPT catalog \citep{ACCEPT_2009}. 
Optically thin cooling is calculated by the exact integration method introduced by \citet{Townsend_2009} and uses solar metallicity cooling tables by \cite{Schure} down to $10^{4.2}$ K. A temperature floor of $10^{4}$ K was also enforced. Purely rotational velocity and magnetic perturbations are seeded at $t=0$ to break the symmetry of the system, with respective dispersion of $\delta v = 75 \, \mathrm{km}\,\mathrm{s}^{-1}$ and $\delta B=1 \, \mu\mathrm{G}$. These perturbations are generated using an inverse parabolic spectrum with a peak scale of 100 kpc. AGN feedback is modeled via source terms for thermal, kinetic and magnetic energy. Cold material (defined as gas cells of temperature $T \leq 5 \times 10^4$ K) is accreted at a rate $\dot{M}_{\mathrm{acc}}$ related to the accretion power by $\dot{E}_{\mathrm{acc}} = \eta \dot{M}_{\mathrm{acc}} c^2$, where $\eta=0.001$ is the accretion efficiency and $c$ the speed of light. This accretion power is then split into the three feedback channels by their corresponding feedback fractions $f_i$, where $i$ designates the kinetic, thermal and magnetic feedback channels. In the absence of star particles, stellar feedback is calculated analytically, by extracting gas denser than a density threshold $n_{\mathrm{thresh.,\star}} = 50 \, \mathrm{cm}^{-3}$ and redistributing it into thermal energy with an efficiency of $5\times10^{-6}$. The structure of the grid is identical in all our runs and kept constant across time. The root mesh grid is made of $1{,}024^3$ cells covering a total volume of $(6.4$\,Mpc$)^3$. The root grid is refined a further $\ell_{\mathrm{max}} = 6$ cubic and nested refinement levels. As a result, the maximally refined region has a volume of $(250 \mathrm{\, kpc})^3$ and is covered by $2,560^3$ cells of size $\Delta_x = 97.7$ pc. The entire analysis presented in this paper is based on data extracted from this maximally refined region, ie. for gas cells located at radii $r \leq 125$ kpc. We use periodic boundary conditions. Considering the total width of the simulated volume, we can exclude the possibility that re-entering sound or shock waves impact our analysis, entirely performed in the inner (200 kpc)$^3$ of the box.

In this study we analyze the data from the fiducial simulation of the XMAGNET suite. This run comprises magnetic fields as well as solar metallicity cooling. The simulation is run for a total duration of 4 Gyr. A detailed analysis of the cluster's evolution in this simulation is presented in \citet{xmagnet_overview}. We note that the feedback power is rather constant for times $t\geq 500$ Myr, with no clear duty cycle. The total cold gas mass varies between $10^{9}$ and $10^{11} \ \rm{M}_\odot$. A visualization of the inner 80 kpc at $t=3.2$ Gyr is presented in Fig.~\ref{fig:Li2020}. It shows a projected map of the cold gas line-of-sight velocity, weighted by H$\alpha$ emission, overlaid on top of a projected temperature map of the hot ICM. A zoom on the innermost region shows a comparison between the original projection, and a smoothed version produced to mimic the effects of atmospheric seeing. 

\subsection{Velocity structure functions}
\label{sect:vsf}
A useful tool to quantify the scaling of turbulence is the $p$th-order VSF. It is a two-point correlation function defined as

\begin{equation}
    S_p(\ell) = \langle \vert\mathbf{u}(\mathbf{x}+\ell\cdot\mathbf{e}) - \mathbf{u}(\mathbf{x})\vert^p \rangle ,
\end{equation}

where $\mathbf{u}$ is the velocity field evaluated at a pair of points of positions $\mathbf{x}$ and $\mathbf{x}+\ell\cdot\mathbf{e}$, and $\mathbf{e}$ is a randomly oriented unit vector. It is expected that the VSF scales as $S_p(\ell) \propto \ell^{p/3}$ in the inertial range, assuming that the turbulence is isotropic, homogeneous and that the fluid is incompressible \citep{Kolmogorov_1941}. To facilitate direct comparison with observations, we focused on the first-order VSFs (i.e., $S_1(\ell)$). It is worth noting that expressing the second-order VSF $S_2(\ell)$ in spectral space leads to the regular Kolmogorov scaling law:

\begin{equation}
    E(k) \propto \varepsilon^{2/3}\,k^{-5/3},
\end{equation}
where $E(k)$ is the turbulent energy spectrum, $\varepsilon$ is the turbulent dissipation rate, and $k$ is the wavenumber. In Appendix~\ref{sect:higherorder}, we also present second-order VSFs for comparison.

We compute the VSF by randomly picking pairs of points separated by a distance $\ell$, calculate their associated velocity difference, and then average them into bins of logarithmically increasing size. Since we compute the VSFs in a region containing a relatively large number of cells (i.e., $1{,}024^3$), we employ a parallelized algorithm adapted from \citet{Federrath_2010}. Sampling pairs of points with separations on the order of a few cell sizes can be inefficient. When randomly selecting two cells in the grid, the probability of choosing a pair with separation $\ell$ is proportional to $\ell^2$. Moreover, since the grid is discretized, the set of possible separations is not continuous. For separations smaller than 500 pc, we adjust the bin sizes to account for the discrete nature of the grid. Specifically, the separation between two cells is given by

\begin{equation}
\label{ijk distance}
\ell = \sqrt{i^2 + j^2 + k^2} \ \Delta_x,
\end{equation}

\begin{figure*}[h!]
    \includegraphics[width=\textwidth]{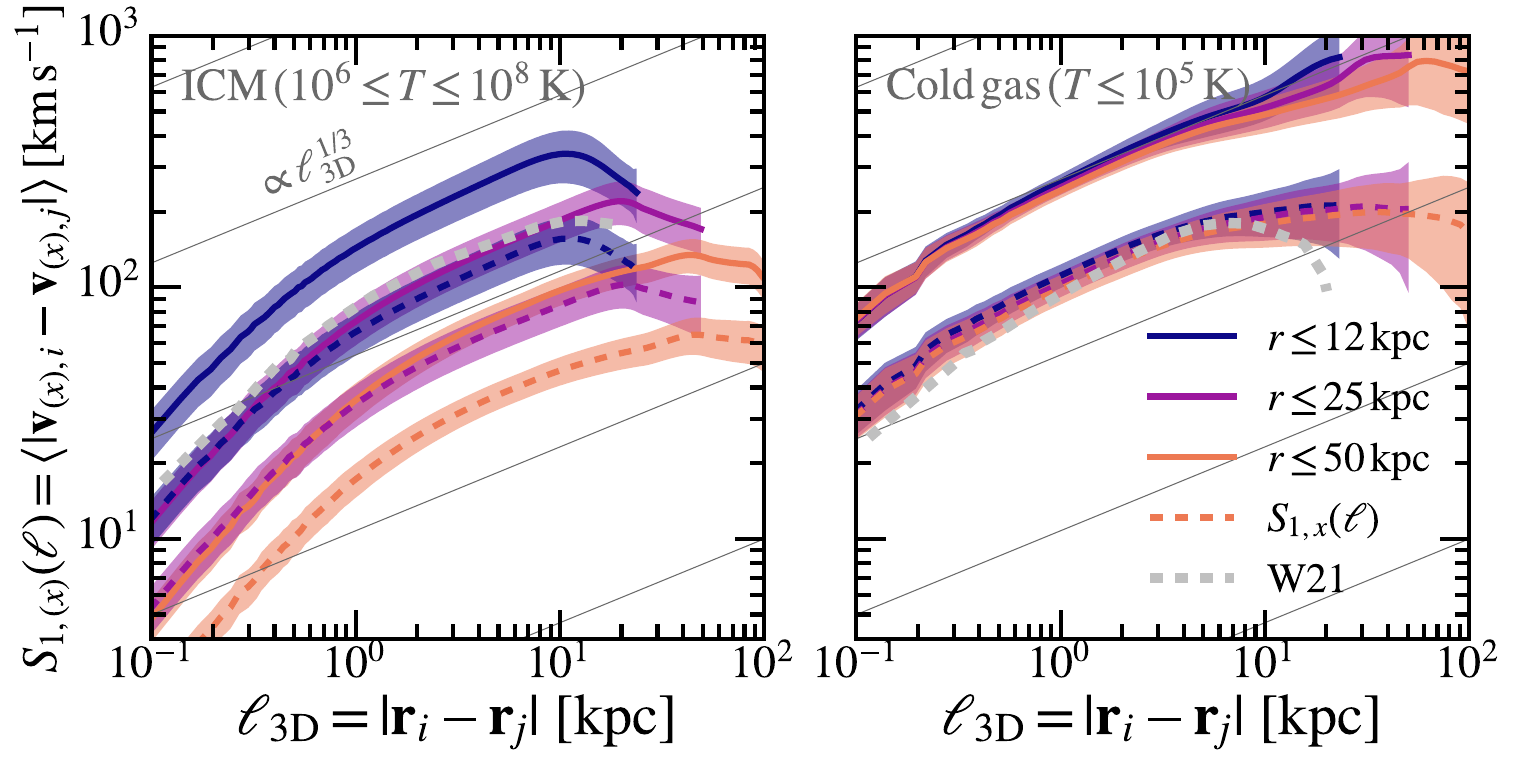}
    \caption{Velocity structure functions of the hot ICM (left panel) and the cold phase (right panel) for all gas located less than 12, 25, and 50 kpc away from the center of the simulated box. The dashed colored lines indicate the line-of-sight VSFs taking the $x$-component of the velocity field. We emphasize that the latter are not equivalent to projected VSFs, as the distance between pairs of cells is still three-dimensional (projected VSFs are discussed in Sect.~\ref{sect:projection}). Colored area is showing standard deviation over time. The solid gray lines in the background indicate the Kolmogorov scaling law \citep{Kolmogorov_1941}. As our setup closely resembles the fiducial MHD run from \citet{Wang_2021}, we present the time-averaged line-of-sight VSFs for the hot and cold phases from that work, for radii $r \lesssim 10$ kpc (dashed gray line labeled W21). A comparison of our results with this study is discussed in Sect.~\ref{sect:comparison_to_studies}. We emphasize that VSFs below separations of $\ell \sim 800$ pc are likely to be steepened by the effect of numerical viscosity.}
    \label{fig:overall_VSF}
\end{figure*}

where $\Delta_x$ is the cell size and $i, j, k$ are integers representing the displacement between the two cells along the three Cartesian directions of the grid. When the grid is two-dimensional (for instance when computing projected VSFs), Eq.~\ref{ijk distance} reduces to $\ell=\sqrt{i^2 + j^2} \ \Delta_x$. To increase the efficiency of the point sampling, we randomly sampled $N$ points on the grid, representing the first pixel of each pair. We then randomly sample a neighboring cell for each of these points following a uniform probability distribution of distance, up to 500 pc. The VSF above 500 pc is computed using a naive sampling of points (i.e., by uniformly sampling both pixels of each pair). The value of 500 pc has been chosen to maximize the efficiency of the code. We have verified that our results are equivalent to the VSF obtained when entirely computed from naive pair sampling. For projected VSF of the cold phase (see Sect. \ref{sect:projection}), the limited number of pixels allowed us to explicitly calculate the velocity difference of all possible pairs of pixels. Convergence tests of our VSF algorithms are presented in Appendix \ref{sect:convergence}.

\section{Quantifying turbulence in the intracluster medium}
\label{sect:quantify}
\subsection{Velocity dispersion}

The time-averaged velocity probability density function (PDF) of the hot and cold gas is presented in Fig.~\ref{fig:velocitydisp}. As the kinematics of the cold phase at the beginning of the active period of the simulation might carry some imprints from the initial conditions, we perform the averaging over all snapshots with time $t\geq 0.9$ Gyr, ie. 400 Myr after the first cold clumps appears. Darker to brighter colored curves represent annuli with increasing distances from the center, namely $r \leq 25$ kpc, $25 \leq r \leq 50$ kpc and $50 \leq r \leq 100$ kpc. The dashed gray curves indicate the data for the gas located at radii $r \leq 100$ kpc. The standard deviation $\sigma(v_x)$ for each curve is derived by fitting a Gaussian function to each distribution. The colored areas represent the standard deviation resulting from time variation. The colored dashed lines indicate the median velocity. The standard deviation of the velocity of the hot gas decreases with increasing radius by roughly a factor of two between the inner and the outer regions of the maximally refined region of our simulated box, with a value of $\sim 60 \, \mathrm{km}\,\mathrm{s}^{-1}$ in the innermost region. This value is relatively low compared to measurements of X-ray calorimeters \citep{Hitomi} of order $\sim 150\,\rm{km}\,\rm{s}^{-1}$, but consistent with other idealized setups similar to ours \citep{Wang_2021,Ehlert_2021}. This could indicate that the contribution of other sources of turbulence, such as mergers, sloshing, or accretion cannot be ignored to reproduce the measured velocity dispersions. We emphasize that our upper temperature cutoff of $10^8$ K excludes the jet material while still including the (shocked) ICM. The velocity distribution of the cold phase is much wider, with dispersions on the order of $\sim 150 \, \mathrm{km}\,\mathrm{s}^{-1}$ in the innermost region. The turbulence within the cold phase is likely supersonic, as the typical velocity fluctuations within the filaments are on the order of $10^2$ km s$^{-1}$, roughly an order of magnitude above the sound speed. We have verified that including emission-weighting does not significantly impact the results for either the hot or the cold phase.

\begin{figure*}[h!]
    \includegraphics[width=\textwidth]{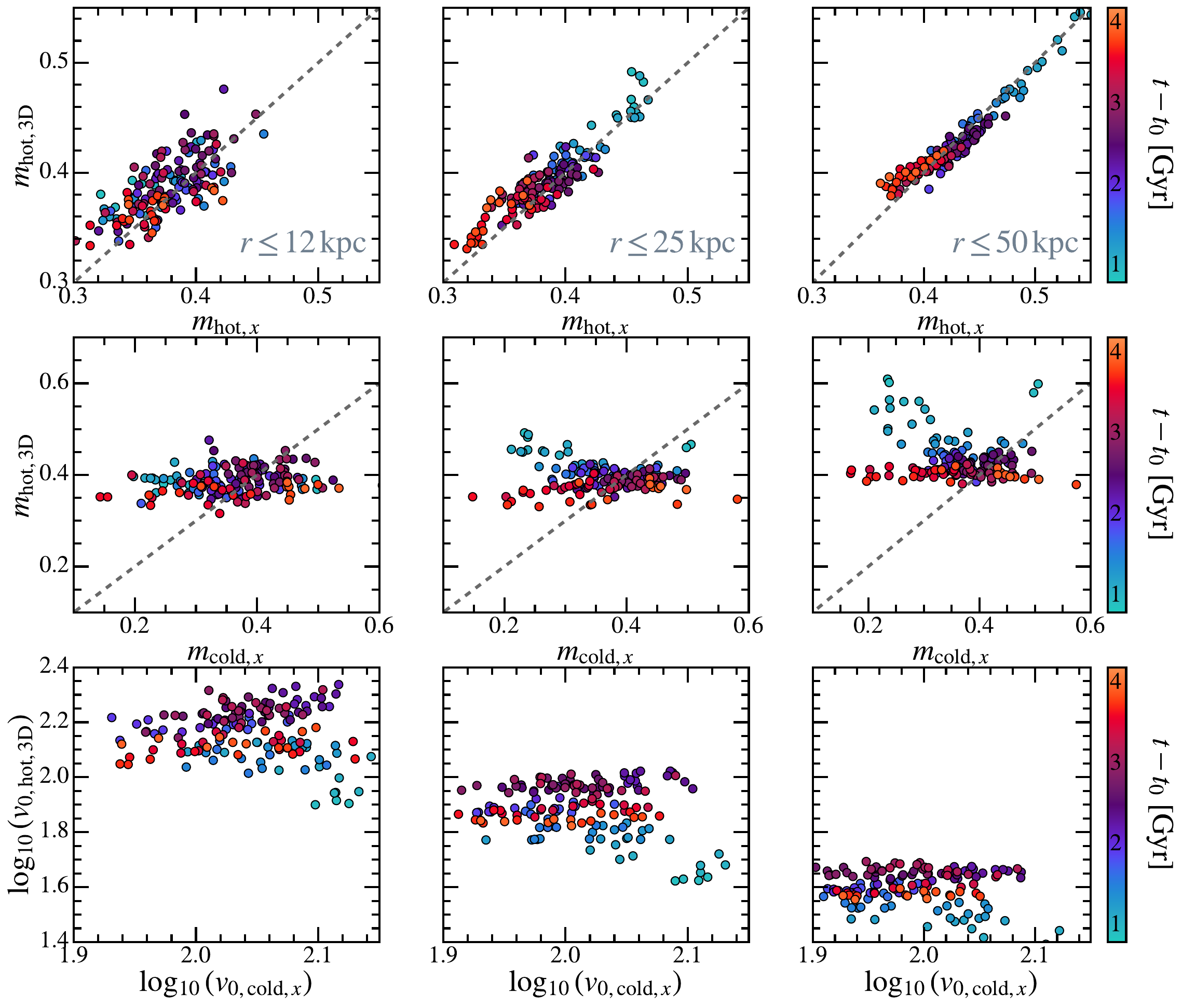}
    \caption{Top panels: Relationship between the power-law index of the line-of-sight hot gas VSF and that of its three-dimensional counterpart. Middle (bottom) panels: Relationship between the power-law index $m$ (amplitude $v_0$) of the line-of-sight, cold phase VSF, and the three-dimensional hot phase VSFm, respectively. The color of the points represents time, with values ranging from 0.9 to 4.0 Gyr. The columns show VSF values for all gas within radii of 12, 25, and 50 kpc (left to right). The dashed gray line indicates the identity relationship and should not be interpreted as an attempt to fit the data. The light blue shaded area represents the region of the parameter space where the measured power-law index of the VSF is steeper than the $m=1/3$ power-law index predicted by the Kolmogorov theory.}
    \label{fig:correlation}
\end{figure*}

\subsection{Radial velocity}

Similarly, we present the distribution of the radial velocity of the hot and cold gas in Fig.~\ref{fig:radveldistrib}. The hot phase (left panel) distribution is characterized by a negative mean radial velocity, as large as $\sim -50 \, \mathrm{km}\,\mathrm{s}^{-1}$ in the innermost region and $\sim -20\, \mathrm{km}\,\mathrm{s}^{-1}$ when including the whole maximally refined region. The velocity distribution of the cold phase is more complex and radially dependent. In the outermost region, it follows a broadly Gaussian distribution with a mean radial velocity of $\sim -400 \, \mathrm{km}\,\mathrm{s}^{-1}$. In the innermost region, the distribution is broader and more asymmetric. At the lower end of the distribution the velocity goes down to $\sim - 10^3 \,\mathrm{km}\,\mathrm{s}^{-1}$. A higher tail of positive radial velocity is also visible and is likely related to uplifting of warm and cold gas resulting from the coupling with the outflowing jets \citep[see also][]{Fournier_2024b}. Overall, the motions of the hot and cold phases are mostly decoupled. The cold phase is dominated by a free-falling motion, while the hot phase only slightly deviates from hydrostatic equilibrium. The resulting turbulent to thermal pressure ratio for the hot phase is typically of $\sim 1\%$.

\begin{figure*}[h!]
    \includegraphics[width=\textwidth]{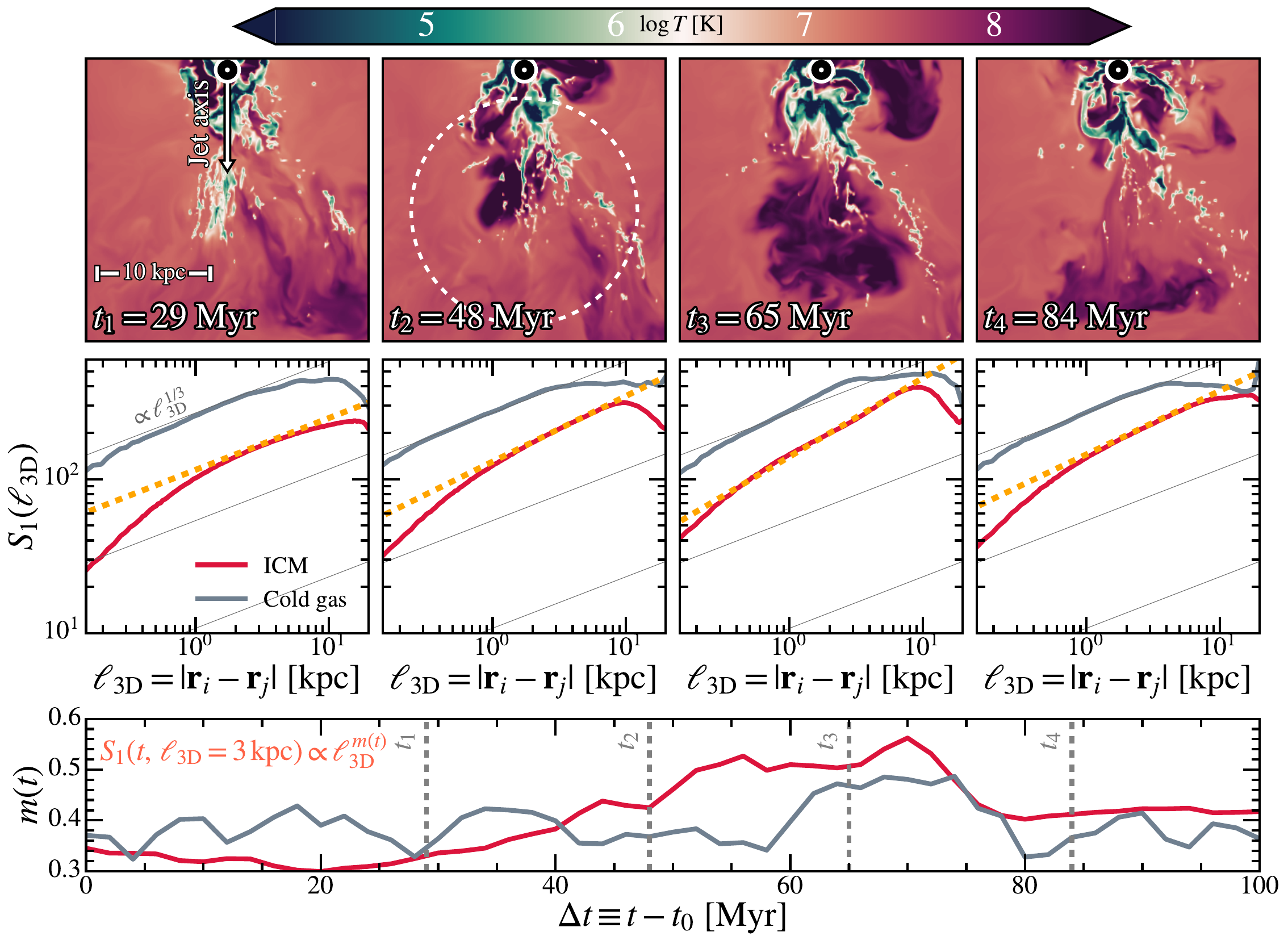}
    \caption{Top panels: temperature projection of the gas. The dashed circle indicates the sphere within which the VSF are computed, whose position is kept fixed throughout the run. The round marker at the top of the temperature projection indicates the position of the SMBH particle and the arrow shows the jet injection axis. The projection depth is 1 kpc. Middle panels: Velocity structure functions of the cold (gray curves) and hot (red curves) phases. The dashed orange lines indicate the fit of the power-law index of the hot phase VSF. Lower panel: Time evolution of the cold and hot phases power-law index. Here, $t_0$ refers to the restarting time of the simulation.}
    \label{fig:cavityVSF}
\end{figure*}

\subsection{Velocity structure functions}
\subsubsection{Overall intracluster medium velocity structure functions}

In Fig.~\ref{fig:overall_VSF}, we present the three-dimensional VSF (i.e., obtained taking the three-dimensional distance between pairs of cells) of the hot  (left panel) and cold (right panel) phases. To ease comparison with observations, we restrict our analysis to spheres of varying radius, centered on the AGN. While a more thorough analysis of individual cold filaments' kinematics might provide deeper insights into their dynamics and evolution, we postpone such analysis to future work. The solid (dashed) lines indicate the VSFs obtained from the magnitude of the three-dimensional velocity vectors ($x$-component of the velocity vectors) of each pairs of cells, respectively. Shaded areas indicate time variation throughout the whole active part of the simulation (i.e., starting from the formation of the first cold structures at $t\sim 0.9$ Gyr). The different colors indicate VSFs obtained including gas located from various radial distances from the center of the box. At low spatial separations (typically up to $\sim$ 1 kpc), the power-law index of the VSFs is likely dominated by effects of numerical viscosity, leading to artificially increasing power-law indices \citep[see also][for a more extensive discussion on this effect]{Grete_2023}. At larger separations, both the time-averaged three-dimensional and line-of-sight VSFs are steeper than the Kolmogorov scaling. There is no significant variations of this scaling when including gas at larger radii. The cold phase VSFs follow a power law with a constant index between $\sim 500$ pc and $\sim 10$ kpc. The time-averaged VSFs are typically steeper than the Kolmogorov scaling index of $m = 1/3$, and on the order of $m\sim 0.4$. In both phases, only including the $x$-component of the velocity vector decreases the amplitude of the VSFs by a factor of $\sim 2-3$, which is larger than the $\sqrt{3}$ ratio expected in the case of isotropic turbulence.
Several previous studies have discussed the possibility of using H$\alpha$-emitting gas or surface brightness fluctuations of the X--ray luminosity as a proxy to trace the hot gas VSF.
In Fig.~\ref{fig:correlation}, we present the relationship between the power-law index of the line-of-sight ($m_{\mathrm{hot},x}$) and unprojected ($m_{\mathrm{hot},3D}$) hot phase VSFs (top panels) and between that of the line-of-sight, cold phase ($m_{\mathrm{cold},x}$) and unprojected hot phase VSFs ($m_{\mathrm{hot},3D}$) (middle panels) for each snapshot. The lower panels present the relationship between the amplitude $v_0$ of the line-of-sight, cold phase VSFs and the unprojected hot phase VSF. The left, center and right panels indicate results for VSFs obtained from all gas cells contained within a limit radius of 12, 25 and 50 kpc, respectively. Each power-law index $m$ and amplitude $v_0$ were measured by fitting the VSFs using

\begin{equation}
    \log_{10}S_1(\ell) = \log_{10}(v_0) + m\log_{10}(\ell),
\end{equation}

where $\ell$ is the separation scale and $v_0$ is defined as the VSF amplitude. The fit is performed between 2 and 4 kpc, avoiding the region influenced by numerical viscosity and staying below the window effect caused by the limited domain size. The gray dashed line indicates the identity relationship.

While the power-law index of the line-of-sight, hot gas VSF shows a good correlation with its three-dimensional counterpart, there is no visible correlation between the power-law index of the line-of-sight, cold gas VSFs and that of the three-dimensional hot gas VSFs. The power-law index of the hot phase VSF is relatively constant throughout the whole duration of the simulation, and is mostly centered around $m\sim 0.4$, with no significant variation when including gas at larger radii (see middle and right columns). The power-law index of the cold phase shows more fluctuations, ranging from $m\sim0.2$ to $m\sim0.5$. The power-law index of the hot phase VSFs are preferentially steeper than their unprojected hot phase counterparts. Similarly, we find no clear correlation between the amplitude of the cold and hot phases VSFs. We emphasize that the cold phase VSFs can vary significantly from one snapshot to an other and thus that many internal details of individual VSFs are smoothed out by time-averaging. Cold phase VSFs at individual snapshots are presented later on (see Sect.~\ref{sect:comparisonwithoptical}).

\subsubsection{Effect of rising cavities}

In \citet{Li_2020} and \citet{Ganguly_2023}, it has been suggested that features observed in the line-of-sight H$\alpha$ emitting gas VSF could be related to the presence and size of the rising cavities related to the AGN activity. In the previous subsection, we have reported the absence of strong correlation between the parameters of the VSFs for the gas located in the inner tens to hundred of kpc of the simulated box. To evaluate the possibility that the AGN activity still might affect the VSF locally, we re-run our cluster setup simulation and save a larger number of snapshots, with an output saving interval of $\Delta t_{\mathrm{snap}} = 2$ Myr, starting from $t = 1.5$ Gyr. 

 In Fig.~\ref{fig:cavityVSF} we present the evolution of the three-dimensional hot phase VSF evaluated in a sphere of radius $r_{\mathrm{sphere}} = 10$ kpc between $t_0=1.4$ Gyr and $t_f = t_0 + 100$ Myr. The sphere was positioned 12 kpc below the SMBH along the z-axis, thus allowing it to encompass the whole volume of a cavity inflated by AGN activity at $t-t_0\sim 65$ Myr. The upper panels of Fig.~\ref{fig:cavityVSF} show projections of the temperature field with a thickness of 1 kpc. The second row shows the corresponding VSFs of the hot (red curves) and cold (gray curves) phases, computed within that sphere. The thin solid gray lines indicate the Kolmogorov scaling law of index $m=1/3$, while the dashed orange line indicates the fit of the hot phase VSFs. In the lower panel, we present the time evolution of the corresponding power-law index for the cold and hot phases. Before the cavity starts forming, the power-law index of the VSF is around $m = 0.3$. As the jets inflate the cavity, the power-law index of the VSF becomes steeper and increases by a factor of $\sim 2$ to reach $\sim 0.5-0.6$. At later times, the cavity rises through the ICM, loses its spherical shape, and exits the sphere of cells under consideration. The power-law index of the VSF decreases again and converges to $m = 0.4$. The VSFs of the cold phase show wider variations and no definitive conclusion can reasonably be drawn from the time evolution of its power-law index. A peak in the VSF at scales of $\sim$ 10 kpc is visible during the phase when the cavity structure is clearly visible.

\section{Effects of projection and comparison to observations}
\label{sect:projection}

\subsection{Mock H$\alpha$ observations}
\label{sect:Halpha}
In \citet{Li_2020} and \citet{Ganguly_2023}, the optical imaging Fourier transform spectrometer (SITELLE) and the Multi Unit Spectroscopic Explorer (MUSE) are used to create maps of the optical emission from the warm ionized phase of several BCGs. The resolved detail in such maps depends on angular resolution and atmospheric seeing, where the latter is standardly modeled by a Gaussian smoothing kernel. For the Perseus cluster, SITELLE's angular resolution corresponds to $\delta x_{\mathrm{pxl}} \sim 250$ pc, and atmospheric seeing induces a FWHM smoothing of $\sim 0.42$ kpc \citep{Li_2020}. To evaluate the effects of spatial resolution and seeing, we produce mock H$\alpha$ emission-weighted velocity maps. We filtered the gas to keep only the cold component of temperature $T\leq 10^5$ K. The emissivity $j_{\mathrm{H}\alpha}$ of the gas in each cell is given by
\begin{equation}
    j_{\mathrm{H}\alpha} = 3 \times 10^{-26} \, T_4^{-0.942-0.031\ln{(T_4)}} n_e n(H^+) \, \mathrm{erg} \, \mathrm{cm}^3 \, \mathrm{s}^{-1} \, \mathrm{sr}^{-1},
\end{equation}
where $T_4 = T / 10^4 \, \mathrm{K}$ and $T$, $n_e$, and $n(H^+)$ are the temperature, electron density, and ionized hydrogen density of each gas cell, respectively \citep{Dong_2011}. The line-of-sight two-dimensional velocity maps were then obtained by projecting the $x$-component of the velocity field on the $(y,z)$ plane. The VSF of the projected map is then obtained by sampling all available pairs of pixels. The separation between each pair of pixels is computed from their projected, two-dimensional distance. We also compute the VSFs obtained from smoothed projected maps to mimic the effect of atmospheric seeing. These maps are obtained by applying a Gaussian kernel to the original H$\alpha$ map (see Fig.~\ref{fig:Li2020}). In Sect~\ref{sect:projprojcold}, we compute VSFs out of projected velocity maps with degraded resolution. The method employed to generate these resolution-downgraded maps is described in Appendix \ref{sect:coarse_graining}.

\subsection{Mock X--ray observations}

\begin{figure}[h!]
\includegraphics[width=0.5\textwidth]{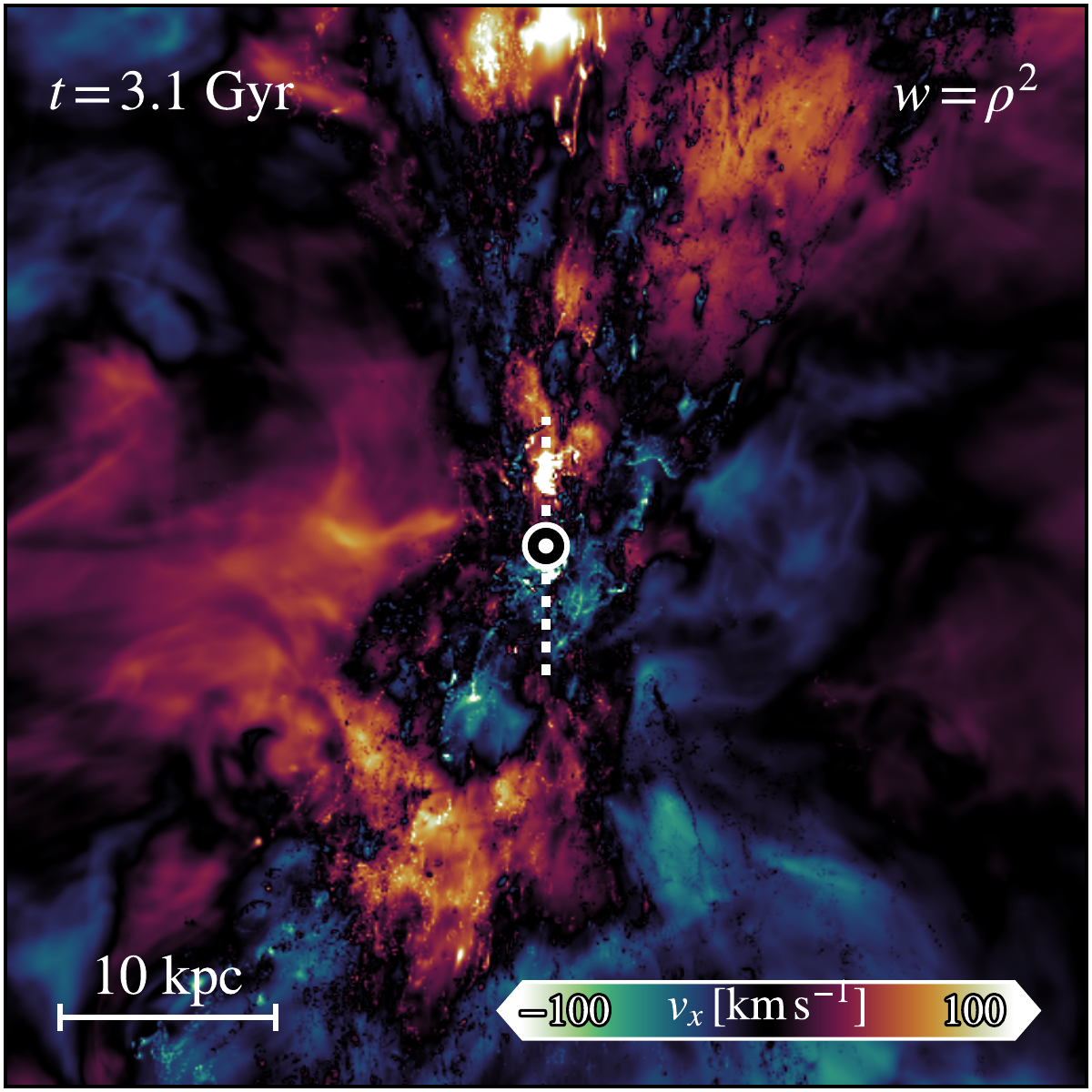}
    \caption{Emission-weighted ($w$) line-of-sight velocity projection of the hot phase within the inner 50 kpc of our fiducial MHD run at $t=3.1$\,Gyr. The marker and dashed line at the center indicate the position of the SMBH particle and the axis of the jets, respectively.}
    \label{fig:HotPhaseProjection}
\end{figure}

We also evaluated the impact of projection on the hot phase.
To this end, we used the same method as for the optical emission. We separated the hot phase by only projecting gas whose temperature is above $10^6$ K and below $10^8$ K (to exclude the jets) and applied either no weighting or a square density weighting. A visualization of such emission-weighted projection is presented in Fig.~\ref{fig:HotPhaseProjection}. Although the cold phase is filtered out from these projections, the continuity of density between the hot ICM and the denser, cold filaments results in over densities in the hot phase enveloping the cold clumps. The square density weighting of the projected map enhance the contribution of these over densities, which are visible in the projected velocity maps. Their impact on the projected hot VSFs is discussed in Sect.~\ref{sect:projectedhotvsf}.

\subsection{Effects of projection and smoothing}
\label{sect:projproj}

\subsubsection{Cold phase}
\label{sect:projprojcold}

\begin{figure}[h!]
\includegraphics[width=0.5\textwidth]{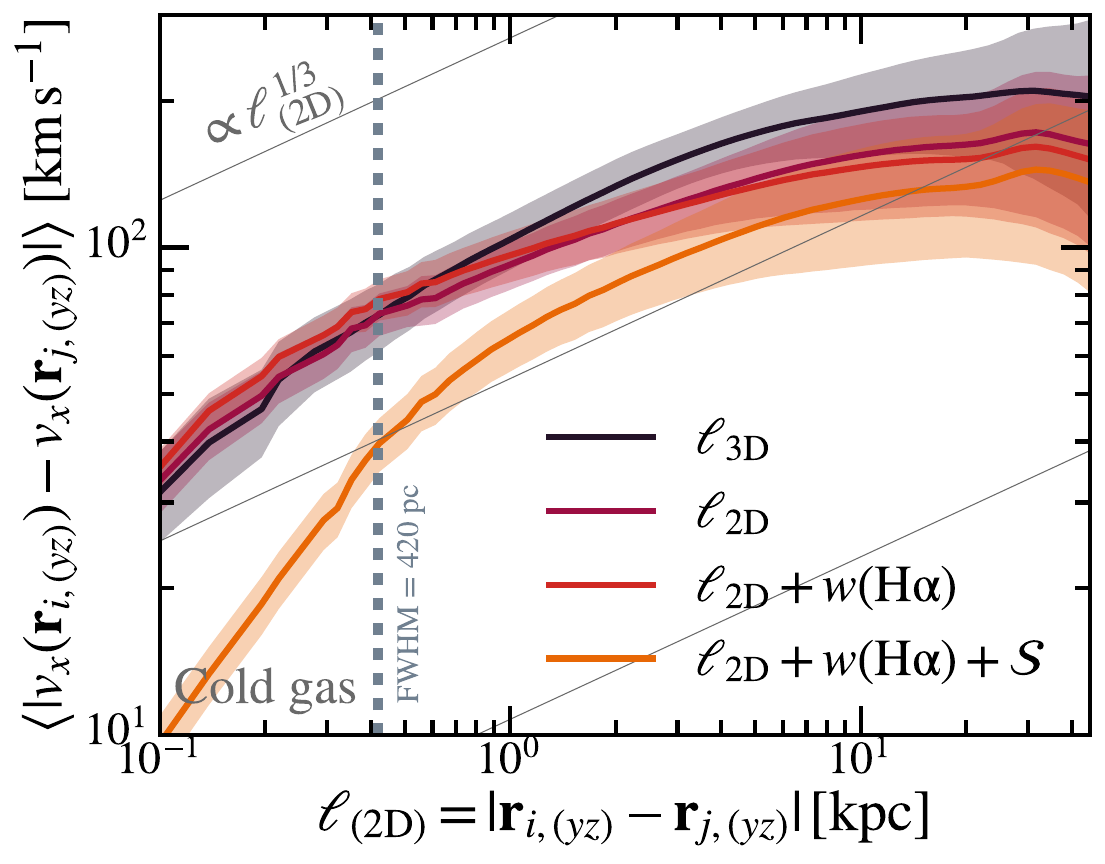}
    \caption{Line-of-sight VSFs for the cold component and for the projected radius $r\leq 25$ kpc averaged between 0.9 and 4 Gyr. The darkest curve (labeled $\ell_\mathrm{3D}$) was obtained by taking the three-dimensional distance between pairs of points, similarly to \citet{Wang_2021}. For all the other curves, the VSFs were computed out of projected maps, and the corresponding separations scales are two dimensional ($\ell_{\mathrm{2D}}$). The term $w(\mathrm{H}\alpha)$ indicates weighting of the map by H$\alpha$ emission, and $\mathcal{S}$ is the additional smoothing effect resulting from atmospheric seeing (see \citet{Li_2020}). The shaded areas indicate the time variation of each VSF.}
    \label{fig:ProjectionEffect}
\end{figure}

\begin{figure}[h!]
\includegraphics[width=0.5\textwidth]{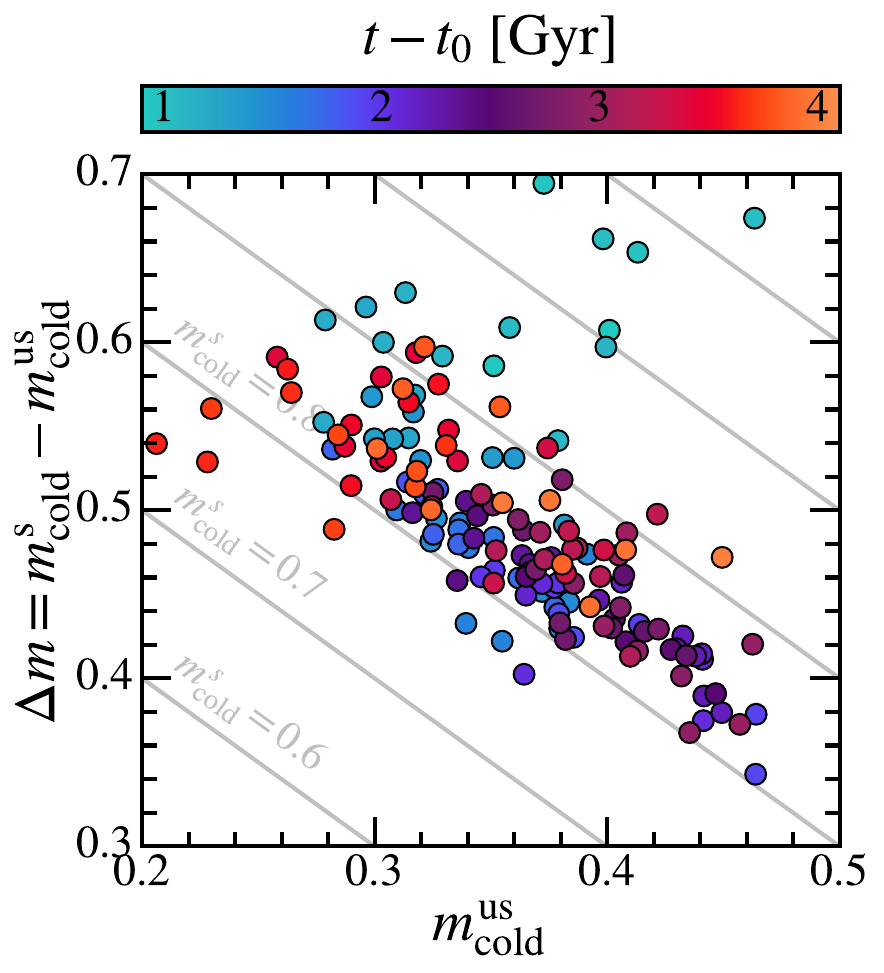}
    \caption{Relationship between the power-law index $m_{\mathrm{cold}}^{\mathrm{us}}$ of the unsmoothed (“us”) projected VSFs (horizontal axis) and $\Delta m$, which represents the increase in the power-law index resulting from smoothing. Lines of equal power-law index for smoothed VSFs ($m_{\mathrm{cold}}^{\mathrm{s}}$) are also indicated for comparison (gray lines).}
    \label{fig:DeltaSlope}
\end{figure}

In Fig.~\ref{fig:ProjectionEffect}, we show the effects of various projection methods on the resulting VSFs. All curves corresponds to the time-average of all VSFs between 0.9 and 4 Gyr for our fiducial MHD run and for all cold gas located at $r \leq 25$ kpc. The colored areas represent the standard deviation resulting from time variation across all snapshots. The darkest curve indicates the VSF of the $x$-component of the velocity field, computed by taking the three-dimensional distance between each pair of cells. All the other curves are computed based on projected maps, implying that separations between pairs of pixels are computed taking their two-dimensional distances. The third curve is computed taking projected maps of the velocity field weighted by H$\alpha$ emission. The last and brightest curve includes smoothing resulting from atmospheric seeing, assuming a smoothing kernel with a standard deviation consistent with the values expected for the Perseus cluster \citep[i.e., a FWHM of 0.42 kpc, see][]{Li_2020}. The blue vertical line indicates the full width at half maximum (FWHM) of the Gaussian kernel used to smooth the projected velocity maps. The effect of smoothing on the VSFs power-law index taking other filters' kernel sizes is discussed in Appendix~\ref{sect:projectiondepth}. 

Taking into account the three-dimensional distribution of the cold gas, the time-averaged VSF is slightly steeper than the Kolmogorov scaling law, with a mean power-law index of $m\sim 0.4$. When calculating the VSF extracted from projected maps, the time-averaged VSF flattens and its power-law index reduces to $m\sim 0.2$. Including weighting by H$\alpha$ emission has little impact on the power-law index, although leading to a minor additional flattening with respect to the unweighted projected VSF. This is consistent with results of previous studies \citep[see, e.g.,][]{Mohapatra_2022}, and is likely related to the low volume filling fraction of the cold phase which results in limited cancellations along the line of sight. Finally, when taking into account the effect of atmospheric seeing on our maps by applying a Gaussian smoothing kernel, the power-law index of the VSF is dramatically increased. Below scales on the order of the smoothing kernel size, the power-law index is close to $m=1$. Above this typical scale, the effect of smoothing on the power-law index becomes less significant, although it still affect the overall amplitude VSF. Since we use a smoothing kernel assuming the distance to the Perseus cluster, the expected effect of smoothing on the line-of-sight velocity maps of galaxy clusters located further away could be more important.

To evaluate the effect of smoothing on individual VSFs for scales on the order of the smoothing kernel size, we present in Fig.~\ref{fig:DeltaSlope} the relationship between the power-law index $m_{\mathrm{cold}}^{\mathrm{us}}$ of the unsmoothed (indicated here and in the figure with “us”) projected VSFs (horizontal axis) and $\Delta m$, which represents the increase in the power-law index resulting from smoothing. All power-law indices are obtained by fitting the VSFs for separations of $\ell = 0.4 \pm 0.1$ kpc. As a comparison, lines of equal smoothed VSFs power-law index $m_{\mathrm{cold}}^{\mathrm{s}}$ are represented with the solid gray lines. Intrinsically shallower VSFs are more affected by smoothing than the steeper ones, as $\Delta m$ varies broadly linearly with $m_{\mathrm{cold}}^{\mathrm{us}}$, with a slope close to --1. Consequently, the power-law index of the smoothed VSFs remains relatively constant across snapshots, with $m_{\mathrm{cold}}^{\mathrm{s}} = 0.86 \pm 0.06$. 

\begin{figure}[h!]
\includegraphics[width=0.5\textwidth]{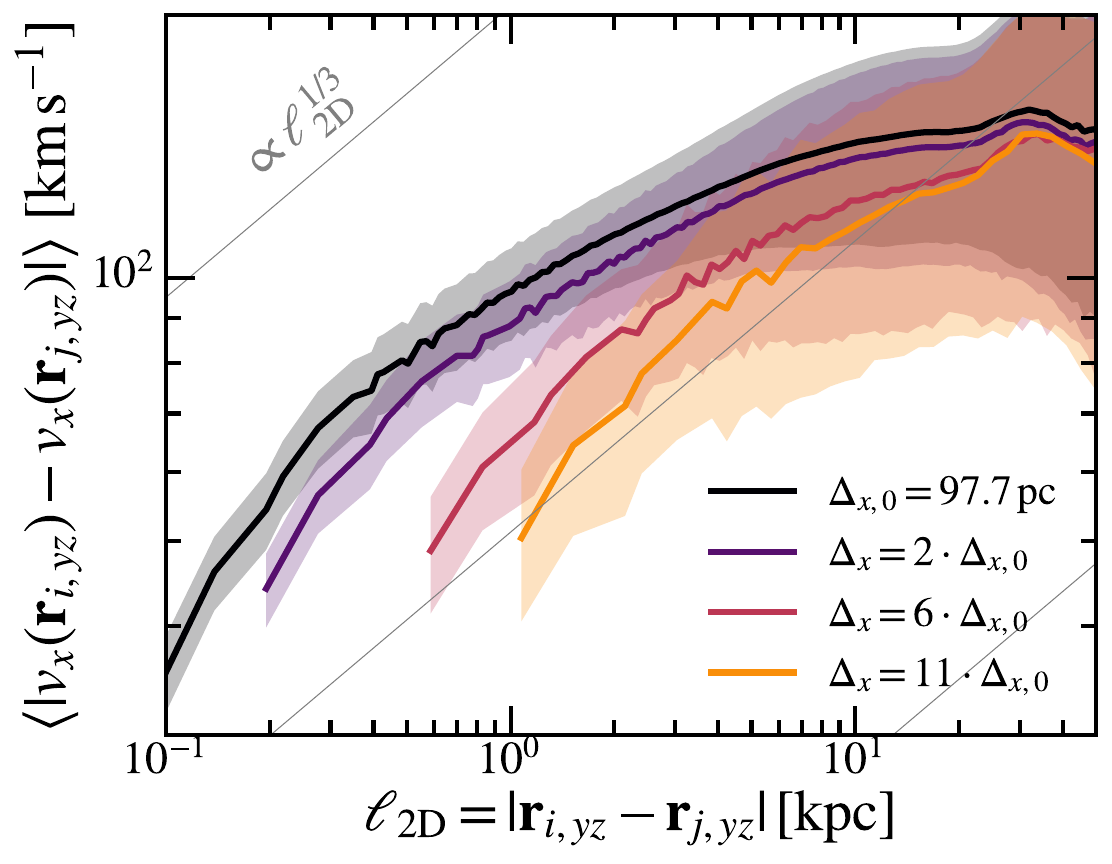}
    \caption{Velocity structure functions obtained from projected maps taking various spatial resolutions for the cold gas located at a projected radius of $r\leq 25$ kpc. The projection method for all curves is equivalent to the third curve in Fig.~\ref{fig:ProjectionEffect} (i.e., labeled as $\ell_{yz} + w(\mathrm{H}\alpha)$). Consequently, the VSFs presented here are shallower than their three-dimensional counterparts (labeled as $\ell_{\mathrm{3D}}$ in Fig.~\ref{fig:ProjectionEffect}) due to projection and emission weighting. The darkest curve corresponds to the maximum spatial resolution $\Delta_{x,0}$ of our simulation. All brighter curves are obtained from coarse-grained maps with pixel size $\Delta_x$. The colored areas represent the standard deviation resulting from time variation across all snapshots.}
    \label{fig:SpatialResolution}
\end{figure}

Instruments used to obtain line-of-sight velocity maps of galaxy clusters have limited angular resolution, translating into limited spatial resolutions depending on the distance of the observed galaxy. To evaluate possible consequences on the VSFs, we present in Fig.~\ref{fig:SpatialResolution} the influence of varying spatial resolution on the velocity maps of the cold phase. While the darkest curve is computed using projected maps with pixels of width $\Delta_{x,0}=97.7$ pc (corresponding to the maximum resolution of our simulation setup), all brighter curves correspond to VSF obtained from maps with degraded resolution, resulting in pixels of width $\Delta_x = k\,\Delta_{x,0}$, where $k$ is an integer. Downgrading the resolution leads to a steepening of the VSFs. The typical VSF amplitude decreases for separations scales lower than the domain size (i.e., for $\ell \sim 25$ kpc). 

\begin{figure}[h!]
\includegraphics[width=0.5\textwidth]{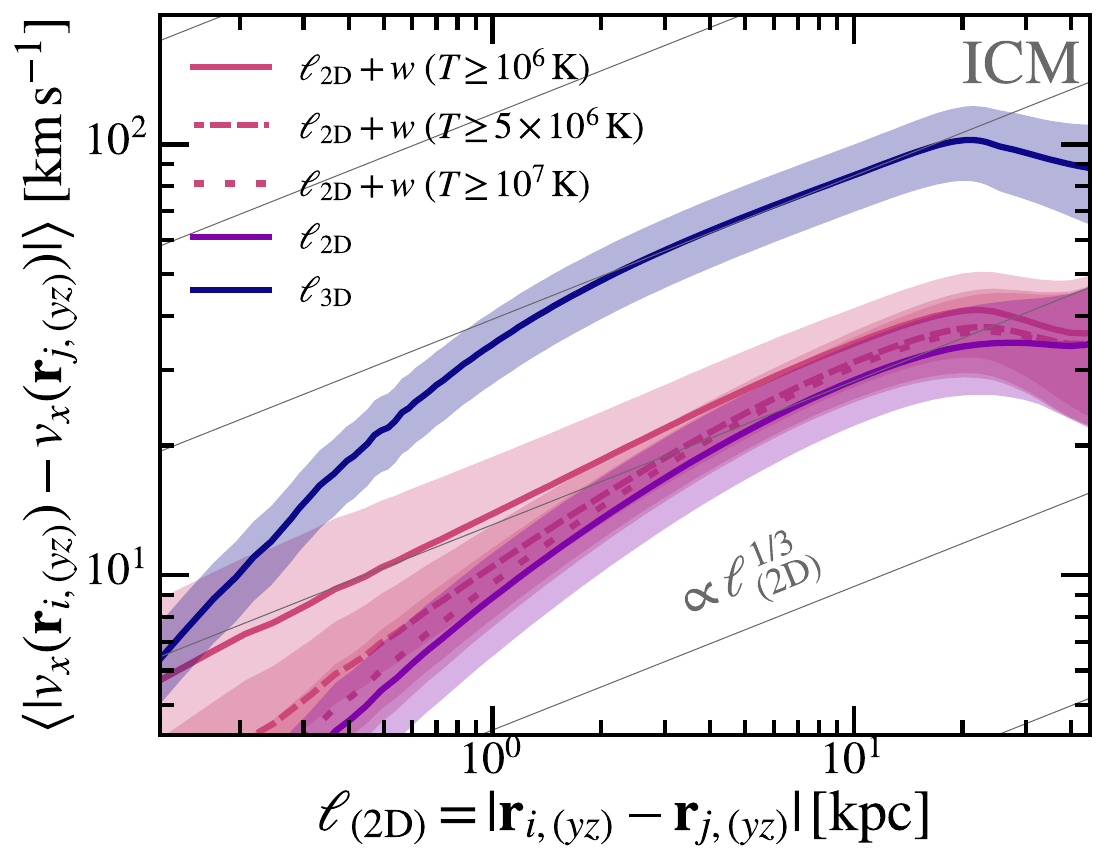}
    \caption{Same as Fig.~\ref{fig:ProjectionEffect} but for the hot phase. The brightest curves indicate the VSF obtained from weighted and projected velocity maps, taking three different minimum temperature cutoffs. The projected VSFs are steeper than the unprojected one for separations above $\ell \sim 1$ kpc.}
    \label{fig:ProjectionEffectHot}
\end{figure}

\subsection{Hot phase}
\label{sect:projectedhotvsf}
In Fig.~\ref{fig:ProjectionEffectHot}, we study the effects of projection on the VSF of the hot phase in a similar way as we did in Fig.~\ref{fig:ProjectionEffect}, but for the cold phase. $w$ indicates that the VSFs were obtained from emission -- weighted velocity maps. The amplitudes of the projected VSFs at any scale are typically lower by a factor of $\sim$2 compared to the three-dimensional VSFs. Both projected VSFs (weighted and unweighted) are steeper than the three-dimensional VSF for separations above $\ell \sim 1$ kpc, and have power-law index of $m\sim1/2$. Including emission weighting makes the VSF shallower than the unweighted ones for lower separations. This is likely resulting from the presence of the over dense hot clumps enveloping the cold filaments (see Fig.~\ref{fig:HotPhaseProjection}). We also note that the steepening of the weighted and projected VSF depends on the choice for the lower temperature cutoff used for the projection. Lower temperature cutoff results in stronger over densities in the hot phase, projected velocity maps with higher contrast, and consequently flatter VSFs. We emphasize that varying the temperature cutoff between $10^6$ and $10^7$ K has no impact on the unprojected or unweighted VSFs. These results are broadly consistent with that reported from turbulent box simulations \citep[see the projected and unprojected VSFs of the radiative MHD run \texttt{f0.10magHR} in Fig. 7 from ][for a direct comparison]{Mohapatra_2022}.

\begin{figure*}[h!]
    \includegraphics[width=\textwidth]{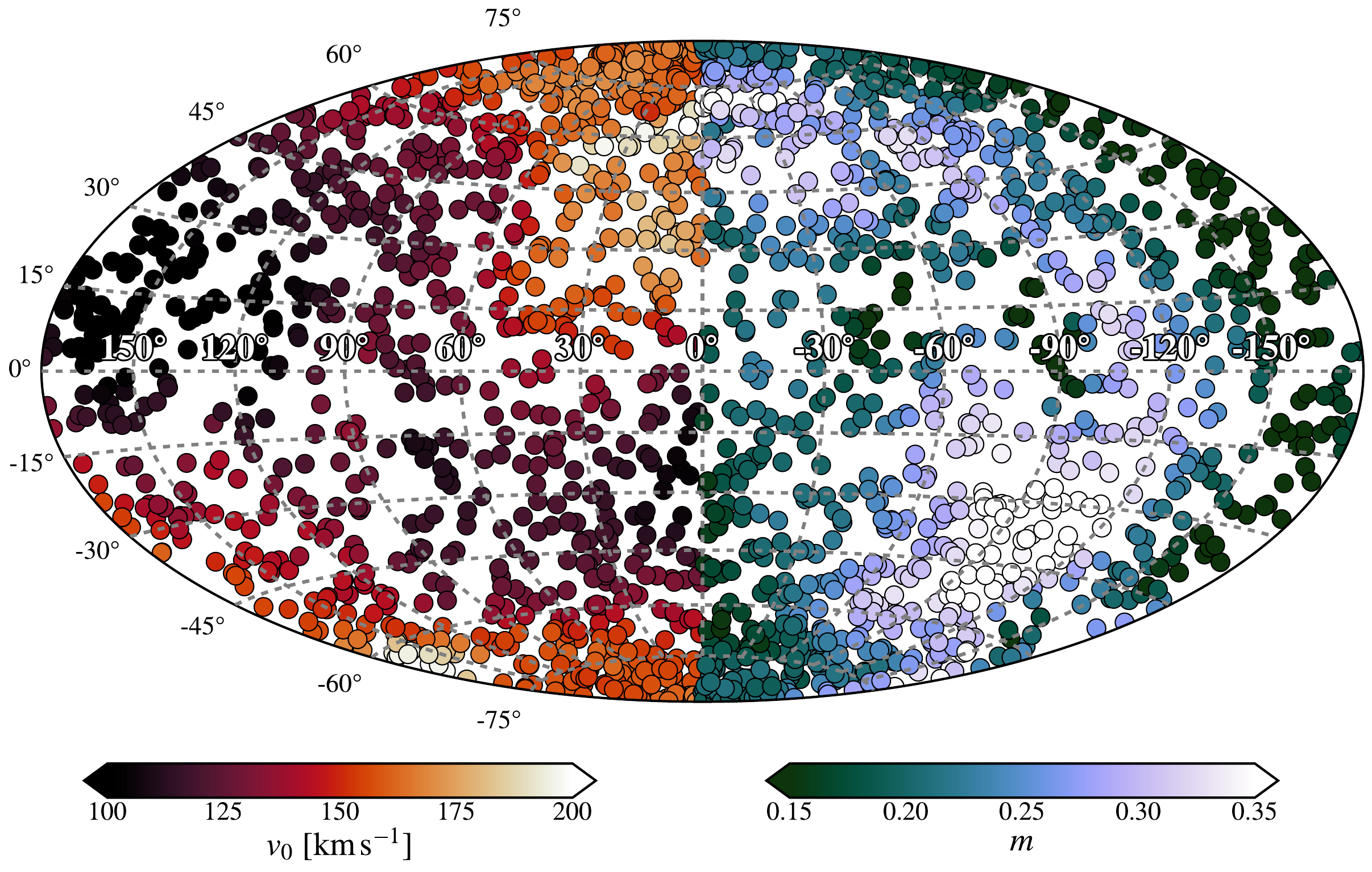}
    \caption{Aitoff projection of the values of the VSF amplitude $v_0$ and the power-law index $m$, measured for 2,000 randomly selected orientations at fixed time $t=3.2$ Gyr within $r=25$~kpc. Each VSF was computed from projected H$\alpha$ emission weighted maps. The power-law fitting was performed between 2 and 4 kpc. As we do not include effects of opacity when generating the projection of the velocity maps used to obtain each VSF, the left and right half of the plot are symmetrical for both $v_0$ and $m$. The coordinates of the jet injection axis is at $\theta = \pm 90^\circ$.}
    \label{fig:allsky}
\end{figure*}

\begin{figure*}[h!]
    \includegraphics[width=\textwidth]{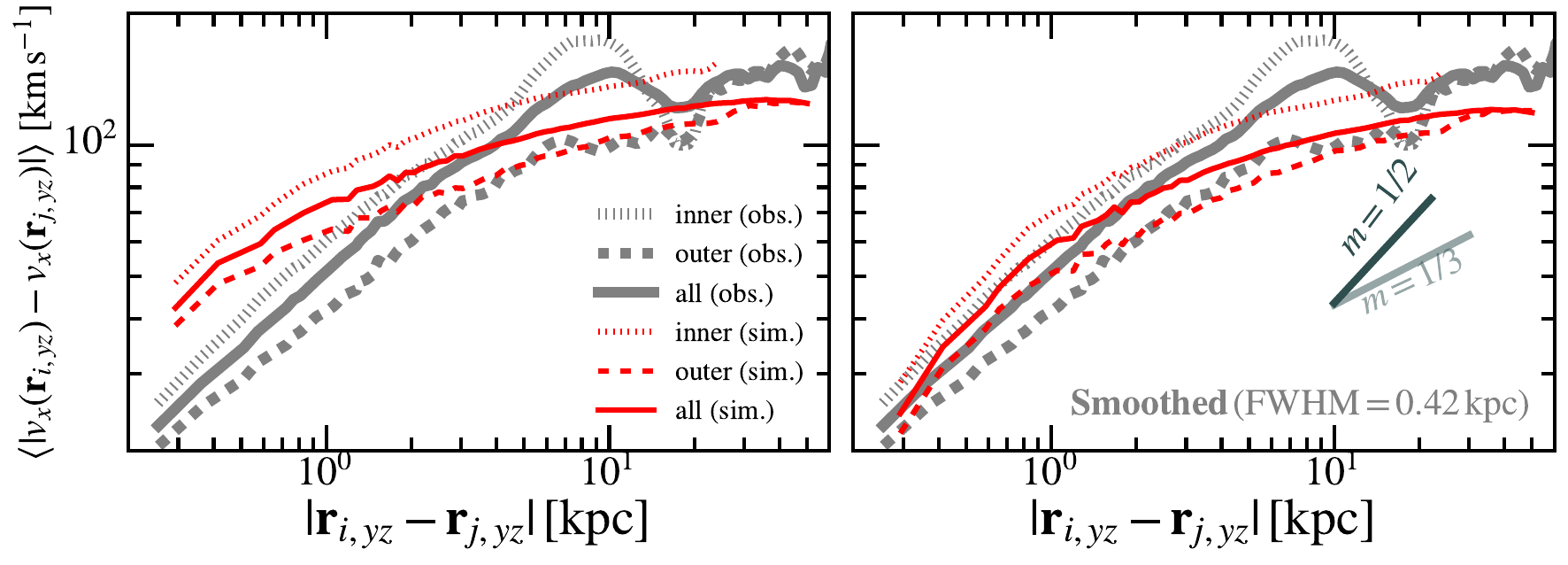}
    \caption{Time-averaged VSFs of the cold phase obtained from mock H$\alpha$ emission-weighted maps projected along the $x$-axis (red lines) compared to observational data \citep[gray lines, from][]{Li_2020}. The images have been coarse-grained to a spatial resolution of 290 pc to mimic the spatial resolution of instruments assuming Perseus cluster distance. The right panel was obtained from smoothed maps to mimic the effects of atmospheric seeing. The labels "inner," "outer," and "all" (see legend) respectively designate the VSFs obtained from all gas cells located at a projected radii of less than 12 kpc, at a projected radius of more than 12 kpc, and for all gas cells irrespective of their radius.}
    \label{fig:Smooth}
\end{figure*}

\subsection{Effects of the observer orientation}

In Fig.~\ref{fig:allsky}, we present a set of 2,000 VSFs computed from H$\alpha$ emission-weighted maps of the cold gas line-of-sight velocity in the inner 50 kpc of our simulated cluster at $t=3.2$ Gyr. Each map is obtained by projecting the line-of-sight velocity of each cold gas cell along a randomly selected orientation identified by its azimuthal and polar angles $(\phi,\,\theta)$. As discussed earlier, the VSF do not follow a power law with a fixed index within the whole interval of separations considered here, as numerical viscosity results in artificially steeper VSF for spatial separations of a few cells. Here, we thus fit the VSF between 2 and 4 kpc, above the region affected by numerical viscosity and below the window effect resulting from the limited domain size. Varying these boundaries does not result in major changes as the power law typically keeps a constant index between 2 and 10 kpc. The values for $v_0$ and $m$ are then presented following an Aitoff projection. Since we do not include effects of opacity during the projection of the velocity map, and the left and right half of the Aitoff projection are symmetrical for both $v_0$ and $m$. The injection axis of the jets corresponds to $\theta = \pm 90^\circ$—that is, the vertical axis at $\phi = 0^\circ$. 

While the typical VSF amplitudes does not vary strongly along the azimuthal angle $\phi$, it varies of nearly a factor of two between viewing angles parallel to the midplane of the simulated box ($\theta = 0^\circ$) and those parallel to the jet axis ($\theta = \pm 90^\circ$). This is likely related to the orientation of the filaments, which tend to align with the AGN jet axis. Since they are preferentially infalling structures, their line-of-sight velocity amplitude is larger along the $z$ axis than the perpendicular directions with respect to the jets. As roughly half of the gas is falling along the $+z$ direction and the other half along the $-z$ direction, this bipolarity results in larger velocity differences at separations on the order $\ell \leq 10$ kpc. The power-law index of the VSF also varies by more than a factor of 2 depending on the observer's orientation. Its value is typically found between $m=0.15$ and $m=0.35$, and is on average shallower than the Kolmogorov value of $1/3$. We attribute this flattening to projection (see Sect.~\ref{sect:projproj}). The polar and azimuthal dependence of the power-law index is likely influenced by the three dimensional orientation of the filaments, and is likely to vary from one snapshot to an other.

\subsection{Comparison with optical observations}
\label{sect:comparisonwithoptical}
In order to compare our results to those in \citet{Li_2020}, we show in Fig.~\ref{fig:Smooth} time-averaged VSFs of the cold gas obtained from H$\alpha$ emission-weighted maps. The maps have been coarse-grained to a spatial resolution of 291 pc (the actual resolution of the SITELLE instrument is approximately 255 pc, but the coarse-graining method used here only permits the resolution to be reduced by integer multiples of the original map's spatial resolution, namely 97 pc).  The labels ``inner,'' ``outer,'' and ``all'' (see legend) designate the VSFs obtained from all gas cells located at projected radii less than 12 kpc, at projected radii more than 12 kpc, and for all gas cells, respectively.  Gray curves correspond to the VSF inferred from optical observations and are taken from \citet{Li_2020}. The left and right panels show results for VSFs calculated from unsmoothed and smoothed, projected velocity maps, respectively. The unsmoothed, coarse-grained map is typically shallower than the Kolmogorov scaling law $m=1/3$. The amplitude at low spatial separations is also greater than observations by a factor of two. Smoothing results in time-averaged VSFs that have power-law index and amplitude more consistent with observations. We note that numerical viscosity can impact the dynamics of gas for separations lower than $\ell \sim$ 1 kpc (see Sect.~\ref{sect:discussion}).

\begin{figure}[h!]
    \centering
    \includegraphics[width=0.5\textwidth]{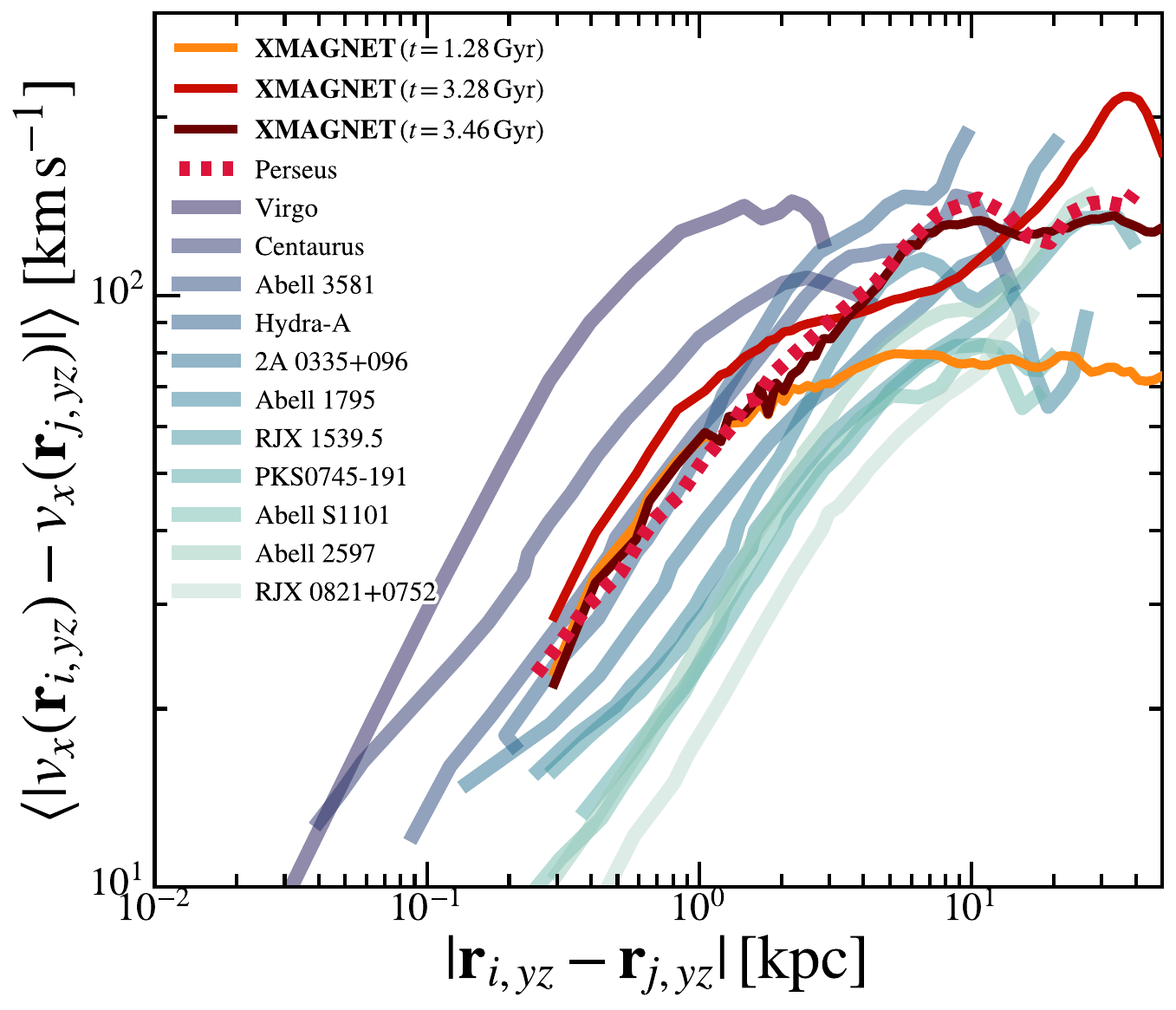}
    \caption{Projected cold gas VSFs including smoothing from atmospheric seeing at three different times in the simulation: 1.28 Gyr, 3.28 Gyr, and 3.48 Gyr (orange, red, and purple lines). For comparison, we also include the 11 VSFs of nearby BCGs from \citet{Ganguly_2023} (thick blue lines and red dashed line).}
    \label{fig:Ganguly}
\end{figure}

Time-averaging VSFs over nearly 4 Gyr (i.e., approximately 170 snapshots) naturally smoothes out many details visible in individual VSFs calculated at fixed times. Although the steepening effect due to smoothing is visible in all VSFs, the relative curvature of the time-averaged smoothed VSFs presented in the right panel of Fig.~\ref{fig:Smooth} does not accurately represent the typical shape of individual VSFs. To provide context for the time-averaged VSFs, we present in Fig.~\ref{fig:Ganguly} individual projected and smoothed VSFs calculated at three different times, namely: 1.28, 3.28, and 3.48 Gyr. For comparison, we also include the 11 VSFs of nearby BCGs from \citet{Ganguly_2023}. Our individual VSFs were selected to highlight typical trends observed throughout our simulation, namely: (i) VSFs characterized by a low amplitude and a relative flatness for separation scales above a few kiloparsecs ($t=1.28$ Gyr), (ii) VSFs exhibiting a distinct peak structure at separation scales beyond 10 kpc ($t=3.28$ Gyr), and (iii) VSFs closely matching the observed VSFs both in terms of steepening and amplitude ($t=3.48$ Gyr). Although a visual inspection of the hot and cold phases suggests that peaks at large separations correlates with the presence and size of X-ray cavities, we were unable to rigorously establish this correlation. This is because the number of pixel pairs at separations approaching the domain size is low, leading to increased statistical uncertainty and significant scatter in the VSF values.

\section{Discussion}
\label{sect:discussion}

\subsection{On the steepening of the velocity structure functions}
\label{sect:steepeningorigin}

Although our results suggest that smoothing from atmospheric seeing may artificially steepen VSFs, they do not rule out the possibility of an intrinsic steepening in either the cold or hot phase VSFs. As shown in Fig.~\ref{fig:DeltaSlope}, the impact of smoothing depends on the intrinsic power-law index of the unsmoothed VSF. Since VSFs with intrinsically steeper slopes experience weaker steepening from atmospheric seeing, it is possible that VSFs measured from observations of nearby BCGs are intrinsically steeper than the Kolmogorov scaling law of $\ell^{1/3}$, artificially steepened by smoothing, or a combination of both. \\
 
Also, several arbitrary parameter choices in our model could significantly impact gas dynamics and, consequently, the resulting VSFs. As discussed in Sect.~\ref{sect:method}, our radiative cooling algorithm assumes a constant and homogeneous metallicity of the gas. This likely leads to an overestimation of the cooling rate at radii beyond $\sim$ 25–50 kpc, which in turn may result in unrealistically extended cold gas structures and eventually excessive radial velocities. Additionally, our feedback parameters—such as jet density—could lead to a different jet-ICM coupling than seen in other simulations. For instance, \citet{Beckmann_2019} demonstrated that precessing low-density jets can strongly interact with the cold phase, producing effects not observed in our setup, such as the coherent outward motion of entire cold gas structures. However, it seems unlikely that the specific choices of our setup alone can fully explain the effects reported in, for example, Fig.\ref{fig:allsky} and \ref{fig:Smooth}, which we attribute to observational limitations. The effect of smoothing on projected maps has been investigated in recent observational studies, which conclude that atmospheric seeing causes a global steepening effect on the VSFs (see notably \citet{Chen_2022} and \cite{LiMcNa_2025}). The absence of magnetic fields in a purely hydrodynamical setup might also lead to steeper VSFs in the cold phase due to the preferentially disky morphologies reported in many previous studies \citep[e.g.,][]{Yuan_2014b,Beckmann_2019,Qiu_2019,Fournier_2024b}. A qualitative comparison between projected cold gas VSFs from the purely hydrodynamical and the fiducial MHD run of the XMAGNET suite is presented in Appendix \ref{sect:hydro_vs_mhd}. We note that recent observational work focusing on jellyfish galaxies has shown that removing rotational components and large-scale motion can lead to a significant reduction in the observed steepness of the VSFs \citep{Chen_2022,Ignesti_2024}. We also emphasize that our results are unlikely to be affected by our choice for the initial velocity perturbations. We have performed several low resolution test simulation and found out that the kinematics of the cold phase is impacted by the initial conditions only during the first  $\sim$ 100 Myr of the active phase (ie. starting 500 Myr after the start of the simulation). We thus intentionally included only data from snapshots with times $t \geq 900$ Myr. It is, however, likely that lower initial magnetic field strengths will result in quantitatively different behavior for the cold phase, closer to purely hydrodynamical setup (see results for the \texttt{weakB} run in \citet{xmagnet_overview}, and associated discussions).\\

Moreover, it has been shown that numerical viscosity affects the motion of the simulated fluid \citep{Grete_2023} and that velocity differences computed between pairs of cells located less than 8--10 cells apart (i.e., up to a physical scale of $\sim1$ kpc in our setup) are thus likely to be underestimated. This is likely to impact the unprojected VSFs presented in Sect.~\ref{sect:quantify} since the separations between pairs of cells are computed in 3D, and we attribute the significant steepening of the hot VSFs at separations lower than 1 kpc to this effect (see Fig.~\ref{fig:overall_VSF}). Evaluating how much this might affect the projected VSFs presented in Sect.~\ref{sect:projection} is not straightforward. Each pixel of the underlying projected velocity maps corresponds to a single velocity value obtained from the averaging of up to hundreds of cells along the line-of-sight, potentially located tens of kpc away from each other. The two-dimensional distance between any given pair of pixels from a projected map is thus not indicative of the underlying three-dimensional distribution of gas cells. This may result in a dilution of the numerical viscosity, and we could expect projected VSFs to be less sensitive to this bias. This effect although remains hard to quantify and we postpone such analysis to future work. \\

\subsection{Comparison to other studies}
\label{sect:comparison_to_studies}

In \citet{Mohapatra_2022}, periodic box simulations with turbulence driving are used to study the effects of radiative cooling, magnetic fields, as well as projection and emission weighting on first- and second-order VSF of the hot and cold phases of the ICM. One key result of their work is that the three-dimensional hot phase VSF is always steeper than the Kolmogorov prediction. Our analysis of the first-order VSF supports such conclusions, as we find a scaling close to $\ell^{1/2}$ near the driving scale, and even steeper at smaller scales (i.e., below $\sim 1$ kpc). However, we find that the cold phase moves differently owing to the radial stratification in the gravitational field of the cluster. Consequently, it is less coupled to the hot phase, and dominated by a preferentially infalling motion toward the center. A direct consequence is that we find the cold phase VSF to be always shallower than the hot phase VSF, and slightly steeper than the Kolmogorov scaling of $\ell^{1/3}$. \\

A simulation setup closer to ours is presented in \citet{Wang_2021}. In this work, an idealized cool-core cluster with AGN feedback, radiative cooling, and magnetic field is followed. Static grid refinement is used and the VSFs are calculated for gas located at distances less than 10 kpc away from the center of the box. The two main differences with respect to our setup are: (i) the grid structure: our maximally refined region has a width of 250 kpc versus $\sim$ 20 kpc in \citet{Wang_2021}, allowing us to quantify turbulence at much larger radii without refinement level transition effects or varied scales of numerical viscosity across levels, and (ii) the AGN feedback mixture: we used a three feedback channels, namely thermal, kinetic, and magnetic, while the AGN feedback is purely kinetic in \citet{Wang_2021}. In this work, the authors conclude that the motion of the cold filaments is mostly driven by gravitational acceleration, possibly slowed down by magnetic fields in the MHD case. The authors also find a correlation between the AGN activity and the power-law index of the hot phase VSF. Quiescent periods give rise to shallower VSFs as compared to active periods. Our results are broadly consistent with the conclusions of this study. The amplitudes and scalings of our line-of-sight VSFs are close to that of their fiducial MHD run (see Fig. 4 in \citet{Wang_2021} and the dashed lines in Fig.~\ref{fig:overall_VSF} of our study for a direct comparison). As reported in \citet{xmagnet_overview}, the AGN activity in our setup does not exhibit significant duty cycles as it is mostly switched on throughout the entire duration of the simulation, with a power of $P_{\mathrm{AGN}} \sim 10^{45} \, \mathrm{erg}\,\mathrm{s}^{-1}$. Thus, we do not observe the same alternation between steeper and flatter VSFs correlated with active and quiescent phases, as shown in Fig. 7 of \citet{Wang_2021}. However, we find evidence that a time-dependent steepening of the VSF is present when the selection of pixel pairs is biased toward regions around rising cavities (see Fig. \ref{fig:cavityVSF}).

\subsection{Missing physics}

Our work presents a simplified and idealized modeling of a cool-core cluster. We have ignored several phenomena and components, which could impact our results. These include the following:

\begin{itemize}
    \item The contribution of orbiting cluster galaxies and sloshing to the local turbulence in the core of the cluster: in our setup, we only model the BCG, although the motion of the other galaxies of the cluster is known to contribute to the overall level of turbulence inside the cluster \citep{Fielding_2020,Ayromlou_2024}.
    
    \item The contribution of cosmic rays, which may affect the self-regulation of cool-core clusters \citep{Jacob_2017,Beckmann_2022} and possibly the formation and morphology of cold structures \citep{Huang_2022}.
    
    \item The impact of a radially dependent metallicity profile: we consider an ICM with a constant metallicity of 1.0 $Z_\odot$. While this reproduces well metallicities measured in the inner tens of kpc of cool-core clusters \citep{Sanders_2007,Mernier_2017,McDonald_2019}, we might overestimate cooling at larger distances (typically at radius $r \geq 25-50$ kpc), resulting in cold structures that are more elongated than measured in real cool-core clusters.
    
    \item Uncertainties on the AGN feedback parameters: many jet parameters remain largely unconstrained, such as the jet material density and temperature. Changing our jet parameters might impact the way AGN outflows propagate \citep[e.g.,][]{Weinberger_2023}, and consequently the kinematics of the cold phase and the development of turbulence \citep[e.g.,][]{Beckmann_2019}.

\end{itemize}

While these simplifications allow us to isolate and study specific physical mechanisms, they also introduce limitations to the general applicability of our results. Future work incorporating these additional processes will be essential to build a more complete and realistic model of cool-core cluster dynamics.

\section{Conclusions}
\label{sect:conclusion}

We have performed MHD simulations of AGN feedback in a cool-core cluster with an unprecedented combination of high resolution ($\Delta_x = 97$ pc) and large grid size ($2{,}560^3$ cells in the inner region). This simulation was run on Frontier, the first exascale supercomputer available to academic researchers~\citep{Atchley2023}, and performed with the AthenaPK code. Our setup includes gravity, radiative cooling, and a model of cold gas-triggered AGN feedback that converts accreted gas into bipolar magnetized kinetic jets and thermal energy. The dimensions and resolution of the innermost maximally refined grid allowed us to study turbulence while minimizing numerical effects from transitions between refinement levels. Our main conclusions are summarized as follows:

\begin{itemize}
        
    \item Based on quantifying the typical line-of-sight and radial velocities of the hot and cold phases, we find that the two phases have significantly different kinematic properties. The hot phase is close to hydrostatic equilibrium, with the velocity dispersions lower than what is measured in clusters with X-ray calorimeters by more than a factor of two. The cold phase follows a ballistic motion and is characterized by infalling and outflowing motions with velocities of up to $\pm 10^3\,\mathrm{km}\,\mathrm{s}^{-1}$.\\

    \item In unprojected VSFs, we find no clear correlation between the VSF properties of the hot and cold phases. The power-law index of the hot phase VSF varies only weakly over time and remains around $m \sim 0.4$. In contrast, the power-law index of the cold-phase VSF fluctuates significantly over time, ranging from $m \sim 0.2$ to $m \sim 0.6$, with a time-averaged value of $m \sim 0.4$.\\
    
    \item We find that the VSF of the hot phase around a rising cavity exhibits a transient steepening, likely related to the injection of turbulence. The power-law index increases by a factor of two, reaching up to $m=0.6$. This steepening lasts for a period of $\sim 50$ Myr, after which the VSF reverts to $m=0.4$.\\
    
    \item Based on our analysis using mock optical images, we conclude that the cold phase VSF is significantly impacted by projection effects. Projection causes substantial flattening of the VSF, and weighting by H$\alpha$ emission further enhances this effect.\\

    \item Gaussian smoothing of the projected maps modeling the effect of atmospheric seeing significantly affects the projected VSFs at any scale. For scales below and on the order of the smoothing kernel size, the VSFs significantly steepen. Intrinsically shallower VSFs are more impacted by smoothing than steeper VSFs. For all separation scales, the amplitude of the VSFs is decreased by a factor of about two.\\
    
    \item Changing the viewing angle leads to a variation of the amplitude and power-law indices of the projected VSFs  by more than a factor of two.\\
    
    \item Projection effects reduce the overall amplitude of the hot-phase VSF compared to its unprojected counterpart and cause a slight steepening.

\end{itemize}

We conclude that projection effects, atmospheric seeing, and the dependence on the viewing angle must be taken into account when interpreting the VSF of the multiphase ICM. Possible extensions of this work include examining the influence of AGN feedback parameters on the VSFs of both the hot and cold phases, exploring potential relationships between VSF features, viewing angle, and the orientation of cold filaments. Incorporating additional sources of turbulence, such as sloshing, would be also be useful to evaluate the relative contribution of the AGN-induced turbulence.

\begin{acknowledgements}
MF thanks Ricarda Beckmann, Tirso Marin-Gilabert and Yuan Li for insightful discussions, as well as Jacob Shen for providing help with visualisation packages. \\
The authors thank Muzi Li and Brian McNamara for sharing a pre-publication draft of their manuscript on velocity structure functions in multiphase galaxy cluster cores.\\
MB and MF acknowledge funding by the Deutsche Forschungsgemeinschaft (DFG, German Research Foundation) under Germany's Excellence Strategy -- EXC 2121 ``Quantum Universe'' --  390833306 and project number 443220636 (DFG research unit FOR 5195: "Relativistic Jets in Active Galaxies").  This research was supported in part by grant NSF PHY-2309135 to the Kavli Institute for Theoretical Physics (KITP).\\
BWO acknowledges support from NSF grants \#1908109 and \#2106575, NASA ATP grants NNX15AP39G and 80NSSC18K1105, and NASA TCAN grant 80NSSC21K1053. \\
This research used resources of the Oak Ridge Leadership Computing Facility at the Oak Ridge National Laboratory, which is supported by the Office of Science of the U.S. Department of Energy under Contract No. DE-AC05-00OR22725. These resources were provided by as part of the DOE INCITE Leadership Computing Program under allocation AST-146 (PI: Brian W. O'Shea).\\
DP is supported by the Royal Society through UKRI grant RF-ERE-210263 (PI: Freeke van de Voort).\\
The authors also gratefully acknowledge the Gauss Centre for Supercomputing e.V. (www.gauss-centre.eu) for funding this project by providing computing time through the John von Neumann Institute for Computing (NIC) on the GCS Supercomputer JUWELS at J\"ulich Supercomputing Centre (JSC). \\
All simulations were performed using the public MHD code {\href{https://github.com/parthenon-hpc-lab/athenapk}{\scshape{AthenaPK}}}, which makes use of the \href{https://github.com/kokkos/kokkos}{{\scshape{Kokkos}}} \citep{Kokkos} library and the \href{https://github.com/parthenon-hpc-lab/parthenon}{{\scshape{Parthenon}}} adaptive mesh refinement framework \citep{Parthenon}. All data analysis was performed with \href{https://yt-project.org/}{{\scshape{yt}}} \citep{Turk_2011,yt4}, \href{https://matplotlib.org/}{{\scshape{Matplotlib}}} \citep{Matplotlib}, \href{https://numpy.org/}{{\scshape{Numpy}}} \citep{Numpy}, \href{https://seaborn.pydata.org/}{{\scshape{Seaborn}}} \citep{Seaborn} and \href{https://cmasher.readthedocs.io/}{{\scshape{CMasher}}} \citep{cmasher}. We thank their authors for making these software and packages publicly available.
\end{acknowledgements}

\bibliographystyle{aa}
\bibliography{biblio}


\begin{appendix}
\onecolumn
\section{Comparison with hydrodynamical setup}
\label{sect:hydro_vs_mhd}

\begin{figure}[h!]
    \centering
    \includegraphics[width=0.5\textwidth]{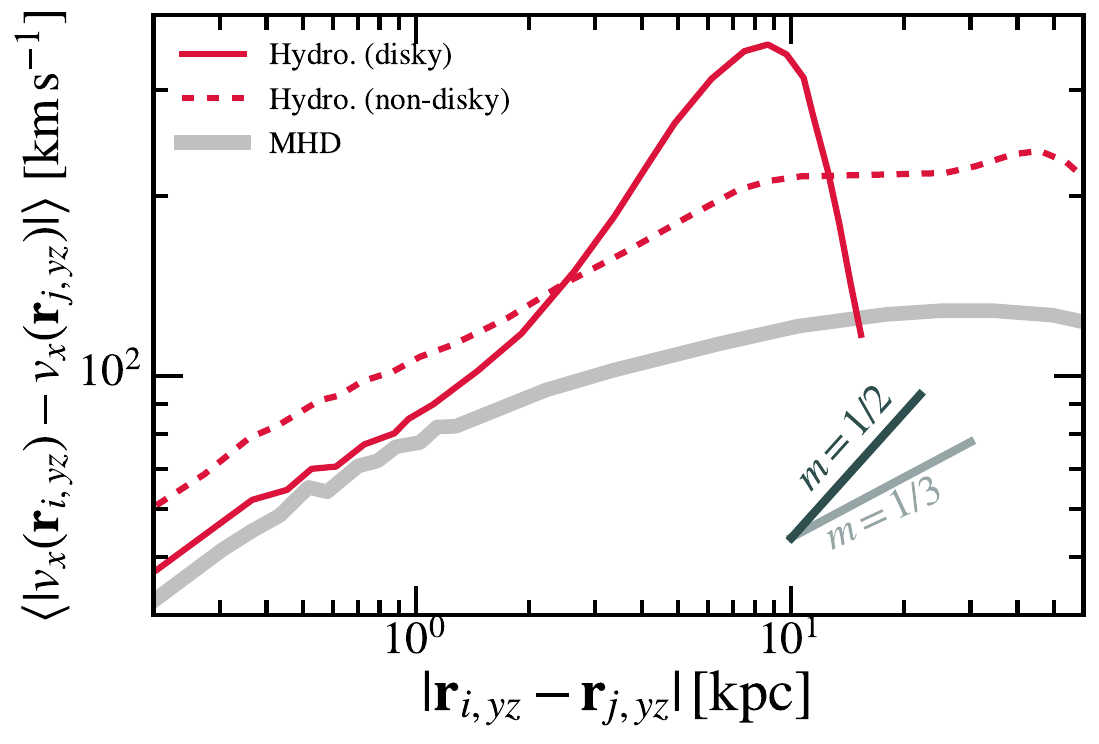}
    \caption{Comparison between the time-averaged projected VSFs of our hydrodynamical (red lines) and MHD (gray line) runs. In the hydrodynamical run, the gas is nearly entirely contained in a disk of radius $r_{\mathrm{disk}}\sim5-10$ kpc for times $t \leq 2.7$ Gyr. We thus distinguish this specific phase (solid line) from the remaining part of the run (dashed line) where no disk is visible (i.e., for $t\geq 2.7$ Gyr).}
    \label{fig:hydro_vs_mhd}
\end{figure}

For brevity, the results section of this paper does not discuss VSFs calculated from the purely hydrodynamical run of the XMAGNET suite. However, it is worth noting that because the structure of the cold phase differs significantly depending on the presence or absence of magnetic fields \citep[see][]{xmagnet_overview}, the resulting VSFs are also markedly different. In Fig.~\ref{fig:hydro_vs_mhd}, we present the time-averaged projected VSFs of the cold gas for our hydrodynamical (red curves) and MHD (gray curve) runs. In the hydrodynamical case, the cold gas resides in a massive disk with a radius of $r_{\mathrm{disk}}\sim 5-10$ kpc, which persists until $t \sim 2.7$ Gyr. After this time, the disk disappears as thousands of cold clumps condenses along the jet axis, increasing the total mass of cold gas in the simulation by nearly two orders of magnitude, up to $\sim 10^{11}\,\mathrm{M}_\odot$ \citep[see Fig. 3 in][]{xmagnet_overview}. The later phase is characterized by ballistic motion closer to what observed in the fiducial MHD run. We distinguish these two phases using two time-averaged VSFs, with the solid line corresponding to the disky phase and the dashed line to the non-disky phase. \\

Because the disk is rotating along an axis broadly aligned with the $z$-axis of the simulation's coordinate system, orthogonal to the projection axis, the resulting projected velocity maps exhibit a distinct pattern where half of the disk has positive line-of-sight velocities and the other half has negative line-of-sight velocities. This results in a characterized peak in the VSFs, visible in the solid line at separation scales broadly consistent with the disk's size (i.e., $r_{\mathrm{disk}}\sim 5-10$ kpc). The time-averaged VSF at later times (i.e., when no disk is visible) does not exhibit any peak. Its scaling is broadly consistent with that of the MHD run (gray curve), although its mean amplitude is larger by a factor of $\sim$2.

\section{Velocity structure function algorithm convergence}
\label{sect:convergence}
\begin{figure}[h!]
    \centering
    \includegraphics[width=0.9\textwidth]{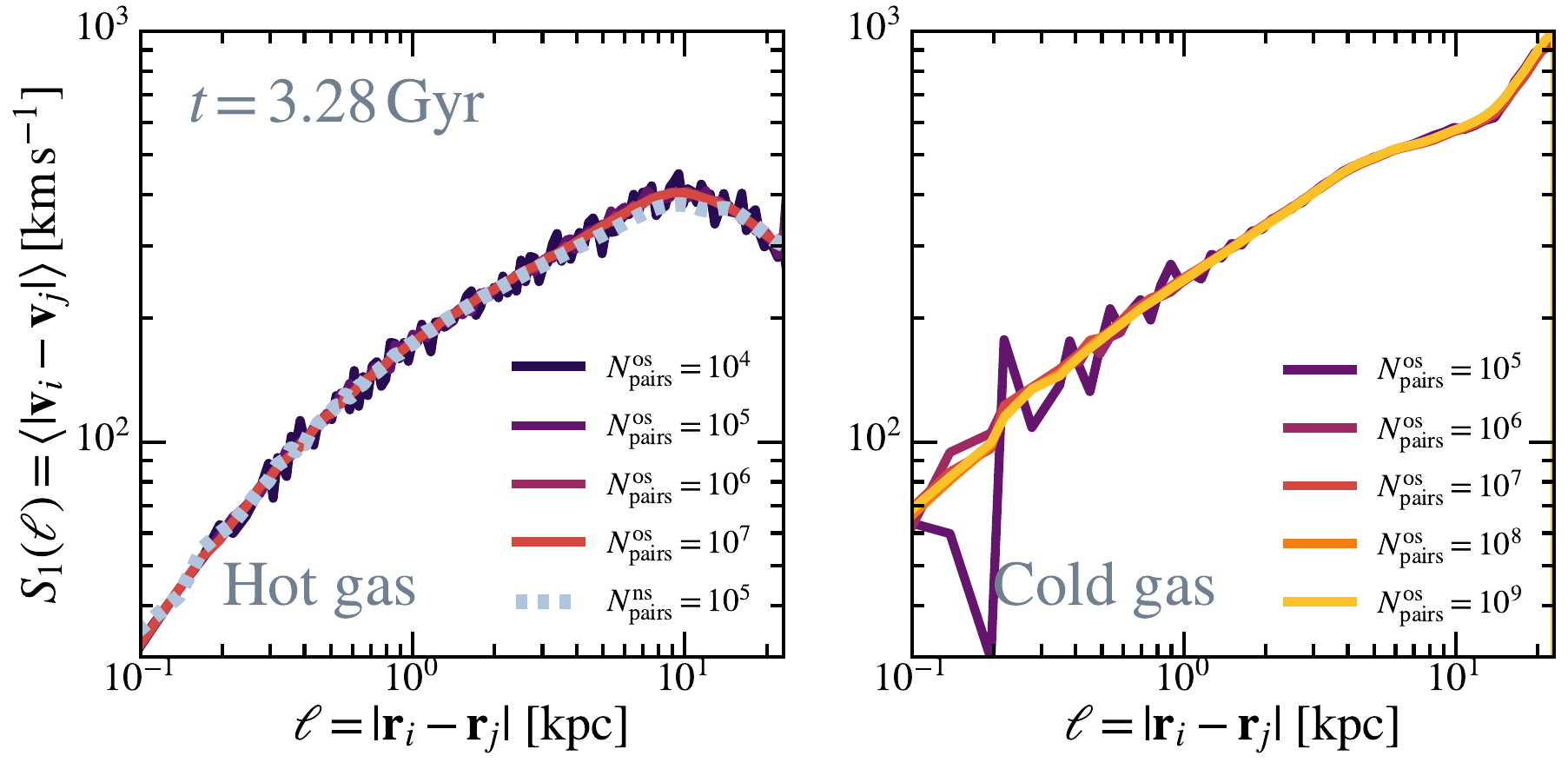}
    \caption{Convergence of the VSF algorithm taking various numbers of pairs of cells for both the hot (left panel) or cold (right panel) phases and for the gas located at radii $r\leq 12$ kpc at $t=3.28$ Gyr. The labels “os” and “ns” designate the optimized and naive sampling methods.}
    \label{fig:Convergence}
\end{figure}

Figure~\ref{fig:Convergence} presents the three-dimensional VSFs of the hot and cold phases at $t=3.28$ Gyr, considering all gas particles within a radius of 12 kpc. For the hot phase, we compare the VSF obtained using a naive sampling method—where all cell pairs are randomly sampled—with our optimized approach, which prioritizes alternative sampling for separations below 500 pc (as detailed in Sect.~\ref{sect:vsf}). As visible, the two methods are in good agreement. We find that VSF convergence is achieved with $10^7$ cell pairs for the hot phase and $10^8$ for the cold phase. However, the volume filling fraction of the cold phase varies significantly throughout the simulation, particularly within the inner 12 kpc, making the number of pairs required for convergence time-dependent. This number also changes with the selected domain size. After testing convergence across multiple snapshots, we adopt $10^7$ ($10^8$) pairs for the hot (cold) phase when analyzing an $r < 12$\,kpc region. For larger domains ($r < 25$ or $r<50$\,kpc), we use $10^8$ ($10^9$) pairs for the hot (cold) phase, respectively.

\section{Comparison with second-order velocity structure functions}
\label{sect:higherorder}

\begin{figure}[h!]
    \centering
    \includegraphics[width=0.9\textwidth]{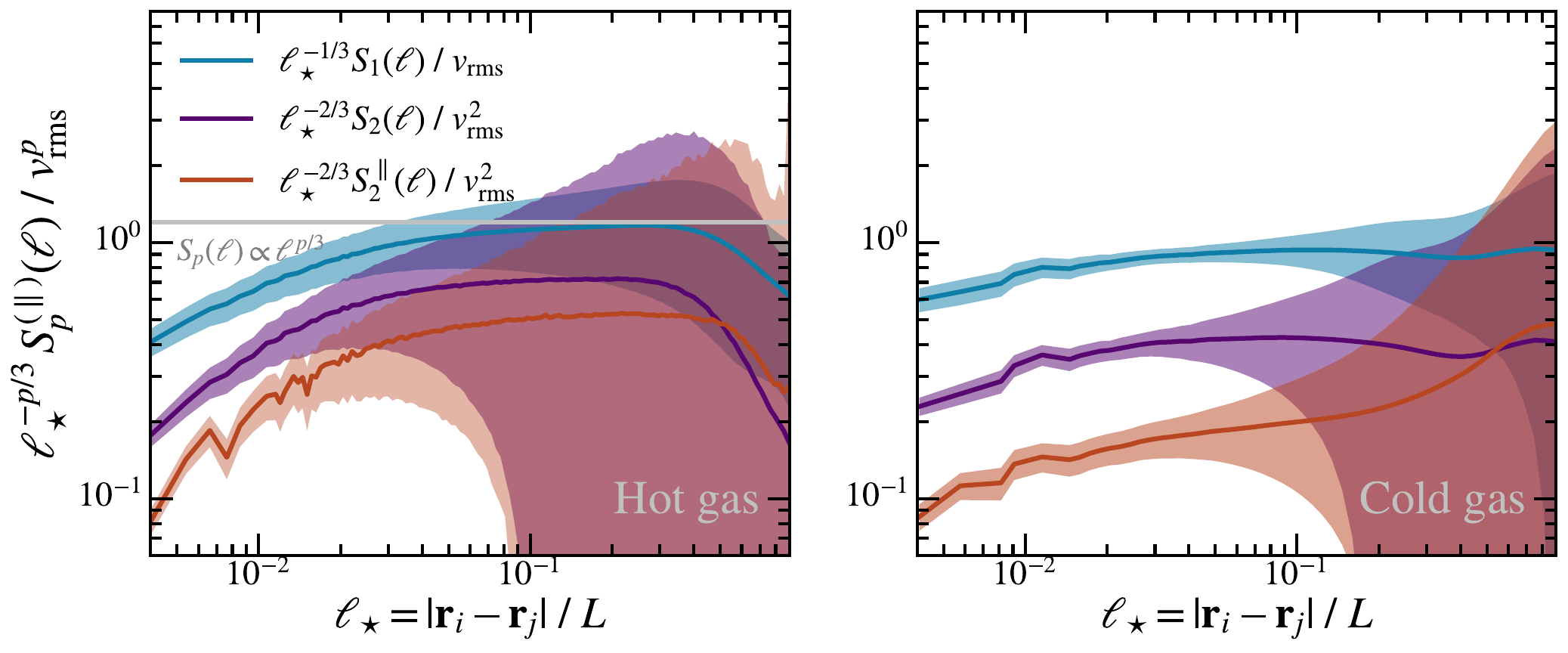}
    \caption{Comparison of the time-averaged first-order VSF (blue curve), second-order VSF (purple curve), and second-order longitudinal VSF (red curve) for the hot phase (left panel) and the cold phase (right panel). The separations scales have been normalized by the domain size $L$, and the VSFs have been normalized by their time-averaged root mean square velocity $v_{\mathrm{rms}}$. The colored areas represent the standard deviation resulting from time variation across all snapshots.}
    \label{fig:Higher_order}
\end{figure}

In Fig.~\ref{fig:Higher_order}, we present a comparison between the first-order (blue curve) and second-order (purple) VSFs for the hot (left panel) and cold (right panel) phases, calculated for the gas cells within $r\leq 12$ kpc. All curves have been rescaled such that a $p$th-order VSF following Kolmogorov scaling, ie. with a power-law index $m=p/3$, appears as a horizontal line in the plot. The Kolmogorov theory of turbulence states that $S_2(\ell)$ and the longitudinal VSF $S_2^\parallel(\ell)$ should scale similarly.  For comparison, we also present the time-averaged second-order longitudinal VSF (red curve), defined as

\begin{equation}
    S^{\parallel}_{2}(\ell) = \langle \{\mathbf{v}(\mathrm{r}_j + \ell \cdot \mathbf{e}) - \mathbf{v}(\mathrm{r}_j)\} \cdot \mathbf{e} \rangle,
\end{equation}

where $\mathbf{e}$ is the unit vector pointing from cell $j$ to cell $i$. The scaling of the first- and second-order VSFs are in good agreement.

\section{Projection depth and smoothing}
\label{sect:projectiondepth}

\begin{figure}[h!]
    \centering
    \includegraphics[width=\textwidth]{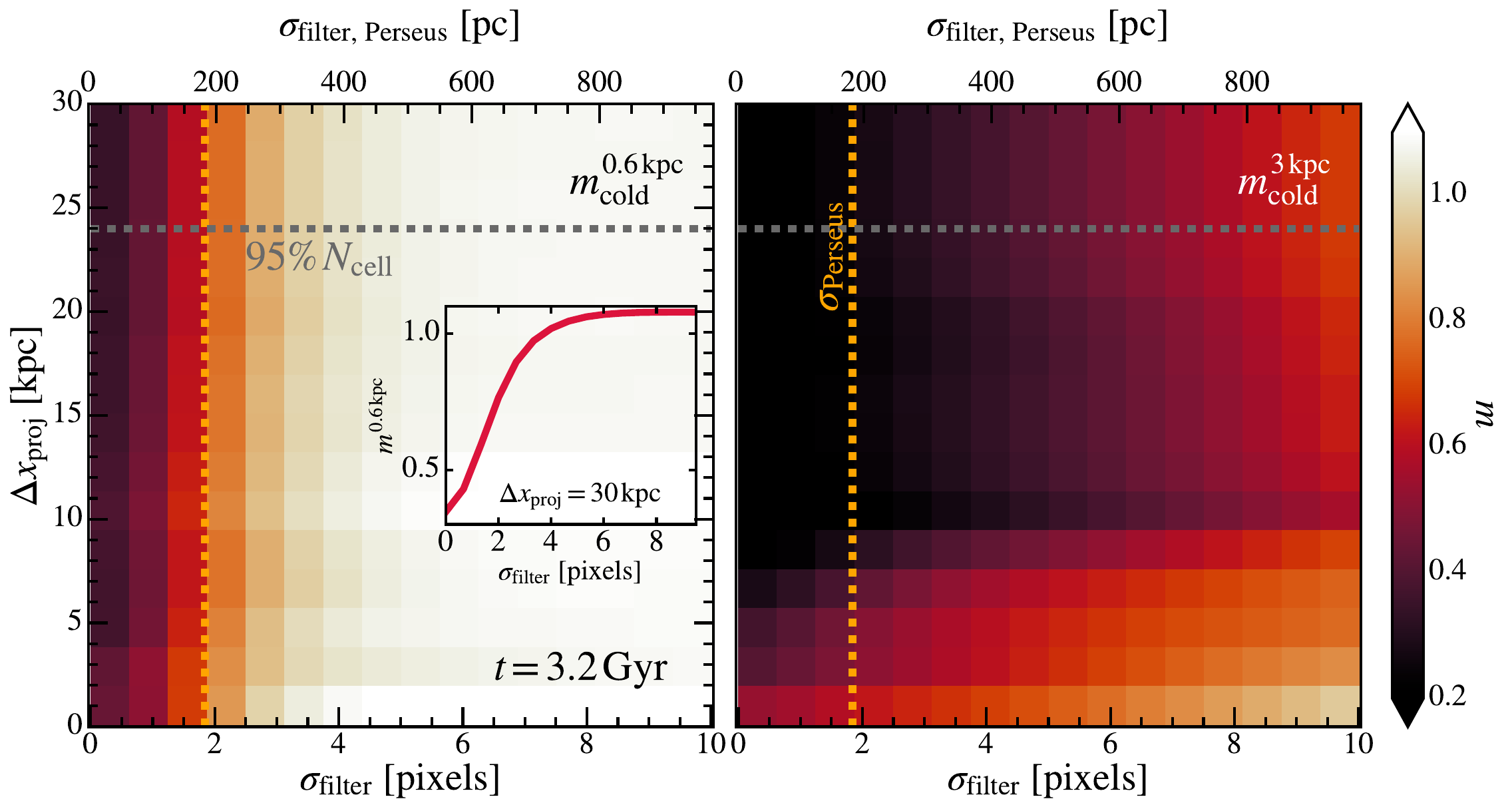}
    \caption{Various power-law indices measured by fitting the VSFs over $\ell = 0.6 \pm 0.1$ kpc (left panel) and $\ell =3 \pm 1$ kpc (right panel) as a function of the Gaussian smoothing kernel size (expressed in pixels in the lower axis and in physical length for the upper axis, assuming the distance to the Perseus cluster for the later) on the horizontal axis and the depth of projection $\Delta x_{\mathrm{proj}}$ along the vertical axis. The vertical orange dashed line indicates the standard deviation $\sigma_{\mathrm{Perseus}}$ of the smoothing effect for the Perseus cluster. The inner figure within the left panel is a 1D visualization of the uppermost row.}
    \label{fig:zproj}
\end{figure}

In Fig.~\ref{fig:zproj}, we present the power-law index of VSFs measured at separation scales of 600 pc (left panel) and 3 kpc (right panel), as a function of the projection depth and smoothing length. Each pixel in the figure represents the power-law index of a VSF computed from a projected velocity map. The projection is smoothed using a 2D Gaussian filter with a standard deviation $\sigma_{\mathrm{filter}}$. The velocity maps include all cold gas cells within a projection depth $\Delta x_{\mathrm{proj}}$, defined by the condition $-\Delta x_{\mathrm{proj}}/2 \leq x \leq \Delta x_{\mathrm{proj}}/2$, where $x$ is the coordinate along the line-of-sight (with the SMBH particle being located at $x_{\mathrm{SMBH}}=0.0$). The horizontal gray dashed line indicates the minimum depth of projection beyond which at least 95 \% of all the gas cells contained in the simulated box are included in the projection. \\

As the projection depth increases more cold structures are included in the projection, modifying the power-law index of the VSFs. This effect is particularly visible in the right panel—that is, for the power-law index measured at 3 kpc. The power-law index measured at 600 pc, the effect of the smoothing is nearly independent of the projection depth, and more significant than at 3 kpc. A smoothing length of 2--3 pixels results in a doubling of the VSFs' power-law index. The inner figure in the left panel is a 1D visualization of the top row (including all cold gas cells of the box), showing the evolution of the VSFs' power law as a function of the smoothing kernel size. For smoothing lengths larger than 4 pixels, the  index converges to $m\sim 1$.

\section{Coarse-grained velocity maps}
\label{sect:coarse_graining}
\begin{figure}[h!]
    \centering
    \includegraphics[width=\textwidth]{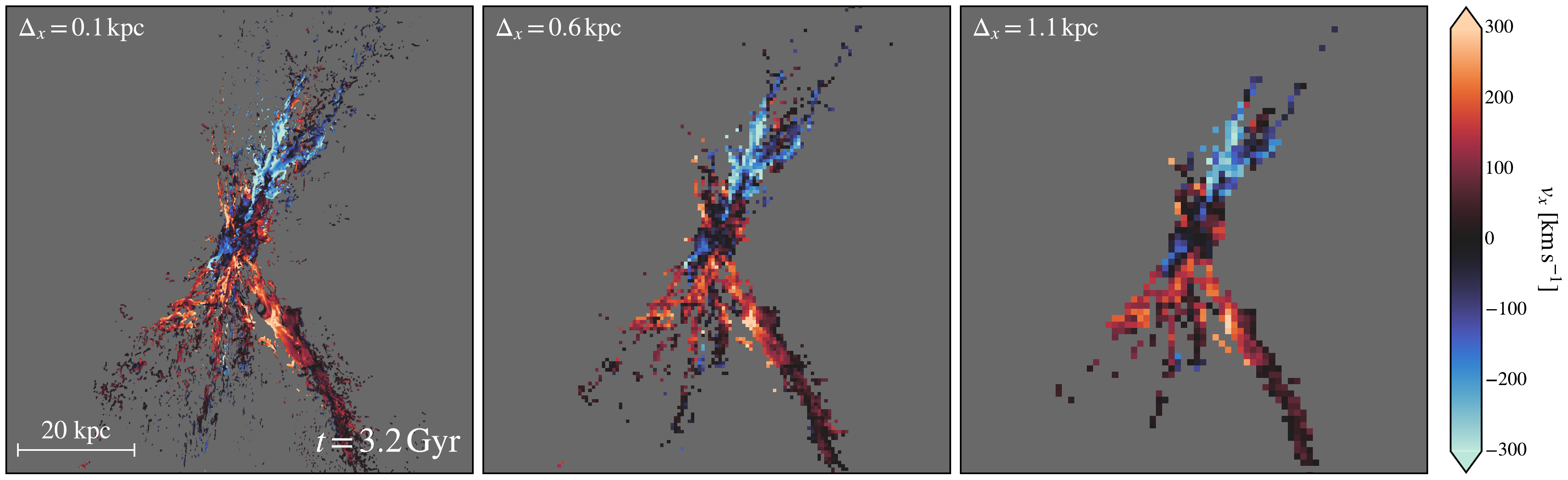}
    \caption{Comparison of three projected line-of-sight velocity maps of the same snapshot and with the same orientation. The panels indicate, from left to right, maps with pixel widths of 0.1 kpc (the maximum resolution of our simulation), 0.6 kpc, and 1.1 kpc.}
    \label{fig:downgrading}
\end{figure}

In Sect.~\ref{sect:projprojcold}, we compute the VSF extracted from projected velocity maps with degraded resolution. To degrade the resolution of the velocity maps, we employ a downsampling technique where the maps are coarsened by a factor \( N \). Specifically, we group adjacent pixels into \( N \times N \) blocks, compute the mean velocity for each block, and assign the result to a single pixel in the downsampled map. During this process, we ensured that blocks with too many empty pixels (i.e., pixels corresponding to the projection of a column containing no cold gas cell) were excluded from the mean computation by applying a threshold $s=0.5$ on the proportion of non-empty pixels per block. A comparison of the resulting maps is presented in Fig.~\ref{fig:downgrading}. This threshold method allows one to ignore small clumps of one or a few cells, which are visible in the high-resolution maps but expected to be spatially unresolved in the degraded map. \\

\end{appendix}

\end{document}